UFMG - ICEx
DEPARTAMENTO DE CIÊNCIA DA COMPUTAÇÃO

UFMG
UNIVERSIDADE FEDERAL DE MINAS GERAIS

# A guide to performing systematic literature reviews in bioinformatics

**Diego C.B. Mariano**
**Carmelina Leite**
**Lucianna H. S. Santos**
**Rafael E. O. Rocha**
**Raquel Cardoso de Melo Minardi**

# A guide to performing systematic literature reviews in bioinformatics

**Diego C. B. Mariano[1]*, Carmelina Leite[1], Lucianna H. S. Santos[1], Rafael E. O. Rocha[1], Raquel C. de Melo-Minardi[1]***

[1] Laboratory of Bioinformatics and Systems (LBS), Department of Computer Science, Federal University of Minas Gerais, CEP 31270-901, Belo Horizonte, Minas Gerais, Brazil

DCBM: diegomariano@ufmg.br, CL: cleite@ufmg.br, LHS: luciannahss@gmail.com, REOR: rafael.eduardo.oliveira.rocha@gmail.com, RCMM: raquelcm@dcc.ufmg.br

***Abstract.*** *Bioinformatics research depends on high-quality databases to provide accurate results. In silico experiments, correctly performed, may prospect novel discoveries and elucidates pathways for biological experiments through data analysis in large scale. However, most biological databases have presented mistakes, such as data incorrectly classified or incomplete information. Also, sometimes, data mining algorithms cannot treat these errors, leading to serious problems for the in silico analysis. Manual curation of data extracted from literature is a possible solution for this problem. Systematic Literature Review (SLR), or Systematic Review, is a method to identify, evaluate and summarize the state-of-the-art of a specific theme. Moreover, SLR allows the collection from databases restrictively, which allows an analysis with lower bias than traditional reviews. The SRL approaches have been widely used for decision-making in medical and environmental studies. However, other research areas, such as bioinformatics, do not have a specific step-by-step to guide researchers undertaking the procedures of an SLR. In this study, we propose a guideline, called BiSRL, to perform SLR in bioinformatics. Our procedures cover the most traditional guides to produce SLRs adapted to bioinformatics. To evaluate our method, we propose a case study to detect and summarize SLRs developed for bioinformatics data. We used two databases: PubMed and ScienceDirect. A total of 207 papers were screened in four steps: title, abstract, diagonal reading, and full-text reading. Four evaluators performed the SLR independently to reduce bias risk. A total of eight papers was included in the SLR case study. The case study demonstrates how to implement the methods of BiSLR to procedure SLR for bioinformatics. BiSLR may guide bioinformaticians to perform systematic reviews reproducible to collect accurate data for higher quality analysis.*

## 1. Introduction

Bioinformatics promotes prospects for novel discoveries and elucidates pathways for biological experiments through data analysis in large scale [1,2]. Research in this area depends on high-quality databases to provide accurate results. However, most biological databases have presented mistakes, such as data incorrectly classified or incomplete information. These errors may be problematic. Current data mining algorithms may be used the filter data. However, occasionally these algorithms cannot treat these errors, leading to serious problems for *in silico* analysis. Manual data curation of data extracted from literature is a solution for this problem. For instance, a database of mutations effects on protein-ligand affinities, composed by 1,000 mutations, was created through a deep search in the literature [3]. Nevertheless, the manual analysis may be subject to bias risk, which may not collect all possible data necessary for a determined study. Manual curation also allows detection of mistakes and, when possible, correct them. Therefore, systematic literature reviews could be used to collect data for bioinformatics analysis.

Systematic Literature Review (SLR), or Systematic Review, is a method to identify, evaluate and summarize the state-of-the-art of a specific theme in the literature. SLR allows the collection of literature information restrictively, that permits a rigorous methodological analysis with lower bias than the traditional reviews [4,5]. In a systematic review, the aim is to construct a general vision of a specific question and give it a fair summary of the literature [5].

To perform systematic reviews is necessary to follow a pre-established and well-defined protocol. Following the defined systematic steps may guarantee the reproducibility of the study. Some guides have been proposed to help reviewers to construct SLRs [5–8], and also evaluation methods of SLR quality, such as the PRISMA statement [9,10]. PRISMA (Preferred Reporting Items for Systematic Reviews and Meta-Analyses) defines a set of items to help authors improve the reporting of SLR. Another example is the Cochrane Collaboration, an international organization that produces systematic literature reviews of interventions in health care, proposing a handbook to help conduct systematic reviews [8]. They recommend that the first step of a systematic review is to develop a protocol that clearly defines the objectives of the review, the criteria for inclusion and exclusion of studies, the methods that identify the studies, and the analysis plan for the collected studies. The main result of a Cochrane Collaboration SLR is a list of the best quality scientific studies for a specific theme. [5,7].

Recently, Pagani *et al.* [7] suggested that the impact factor, number of citations, and year of publication are important to select and rank relevant scientific papers encompassing the reviewed subject. Also, Khan *et al.* [6] defined that an SLR can be performed in five steps: construction of a question for the review; the identification of relevant works; evaluation of the studies collected; summarizing and data synthesis; and interpretation of the findings. However, their step-by-step explanation was developed for healthcare based works.

In the medical field, SLR help to establish evidence based on clinical practice, to improve the effectiveness of interventions, to gather incidence or risk factor for diseases, to diagnostic test accuracy or patient experience [5]. Although SLRs have been widely used for decision-making in medical and environmental research, the principles can be adapted for other fields. Hence, several areas can take advantage of systematic reviews by just performing some adaptations in these methodologies.

Kitchenham *et al.* [4,11] suggested a procedure for performing systematic reviews for software engineering researchers. In their work, a guideline, derived from three existing guidelines used by medical researchers, was proposed. However, adapted to reflect the specific problems of software engineering research. The same strategy recommended by them can be used to create specifics guidelines for new areas, such as bioinformatics.

In this work, we suggest a guide accomplish SLRs in bioinformatics, called BiSLR. A guide to performing systematic reviews in bioinformatics should consider that the potential users are deriving from several areas. Our procedures cover the most traditional guides to produce SLRs and adapt them to bioinformatics. To best of our knowledge, this is the first guideline to perform SLR in bioinformatics. Also, we performed a case study to explain the BiSLR steps better.

## 2. BiSLR

In this section, we describe the steps to perform an SLR in bioinformatics. Then, we proposed a case study to validate our method and also visualize the state-of-art of SLRs in bioinformatics. BiSLR was designed to summarize the state-of-art of a bioinformatics question. In this guide, we also make suggestions to minimize the risk of selection bias that the potential readers, bioinformatics professionals, mainly Ph.D. students, may encounter. We believe that this guide can increase the number of SLRs published in bioinformatics research, and also improve the quality and meaning of the studies in the literature for Ph.D. students.

BiSLR proposes that the process to construct an SLR is iterative. Hence, the steps should be repeated until the data collected to comply with the pre-established goals of the SLR. For this reason, BiSLR proposes a spiral model (Figure 1). During the execution of an SLR, some steps must be repeated until a satisfactory result is obtained. The spiral model demonstrates that the activities flux in the BiSLR's procedures crosses repeatedly a set of steps. BiSLR is divided into four steps: (i) protocol definition: where the objectives and the protocol for performing the SLR are defined (Figure 2a); (ii) reference collection: where the bibliographic references are collected from literature (Figure 2b); (iii) data evaluation: where the references collected are evaluated based on the objectives and questions defined in the protocol (Figure 2c); and (iv) interpretation of the findings: where the findings are summarized and curated biological data is collected (Figure 2d).

### 2.1. Protocol definition

A protocol is a set of restricted definitions that will guide the SLR execution. The protocol registers the SLR main information, such as the main question and specific questions, SLR objectives, criteria for inclusion and exclusion (Table 1).

The main question of a study is the first step of an SLR. This main inquiry will guide the SLR. Its definition is empirical, obtained by empirical knowledge, and it can be performed with the help of interviews with specialists in the research area, general searches on the Internet, or reading relevant literature references in the area. The main question may be a simple and generic question, *e.g.* "what is the most used next-generation sequencing technology for prokaryotic organisms sequencing" or "what are the database management systems recommended for dealing with transcriptome data". We advise that the head of the study selects a group of researchers (postgraduates, undergraduates, scholars, etc.) and perform a brainstorm section to construct the main question. An SRL must not be performed alone. Once the main question is defined, it cannot be changed. The objectives of the research also must be documented. Then, it is necessary to define general criteria for a published study to be included or excluded from the SLR list. Also, five specific questions can be defined. These questions can be slightly changed until the last part of the collected studies evaluation step. Then, include in the SLR a table registering all that information (Table 1).

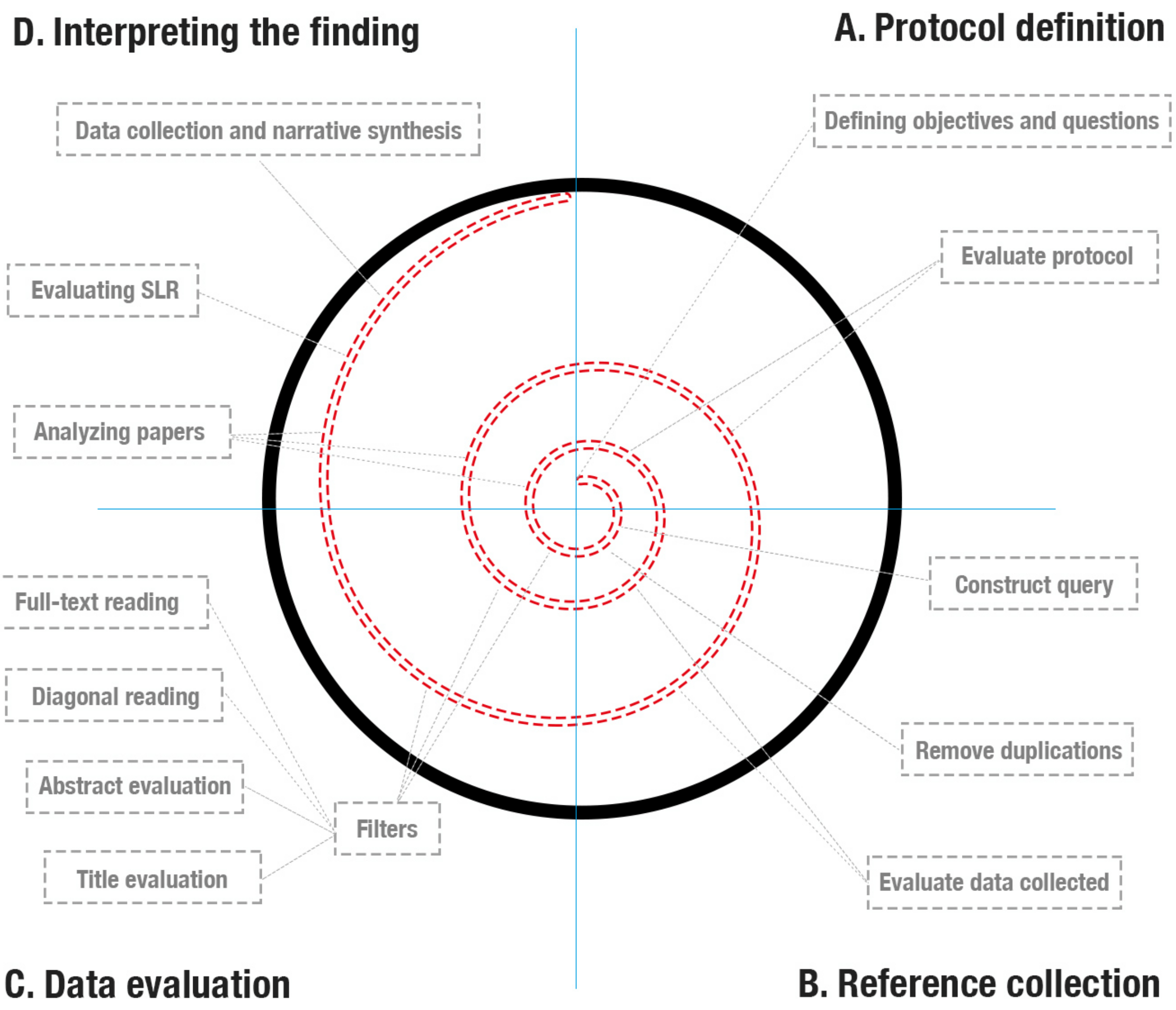

**Figure 1. BiSLR spiral model.**

The execution of an SLR should be performed iteratively. For this reason, BiSLR proposes a spiral model. BiSLR recommends restarting the process if any possible mistake was detected. The BiSLR model is divided into four main steps. (A) Protocol definition: determines the description of the SLR protocol. In this step, the reviewers define objectives and questions that will guide the SLR. (B) Reference collection: define the references search in the literature. (C) Data evaluation: consists in assessing each study collected in the SLR and define the list of papers included in the SLR. (D) Interpreting the findings: consists in analyzing the papers covered in the SLR, group, summarizing, evaluate the SLR, collect data for construction of curated databases and construct a narrative synthesis to summarize everything.

## 2.2. Reference collection

The first step of reference collection is defining databases for searching. Bibliographic databases store publications from specific areas. Hence the database choice depends on the SLR objectives (a list of databases is available at supplementary material). We recommend that more than one database be used in the search step. Redundant data must be removed. We provide a Python script to remove redundant references based on DOI (Digital Object Identifier) at supplementary material.

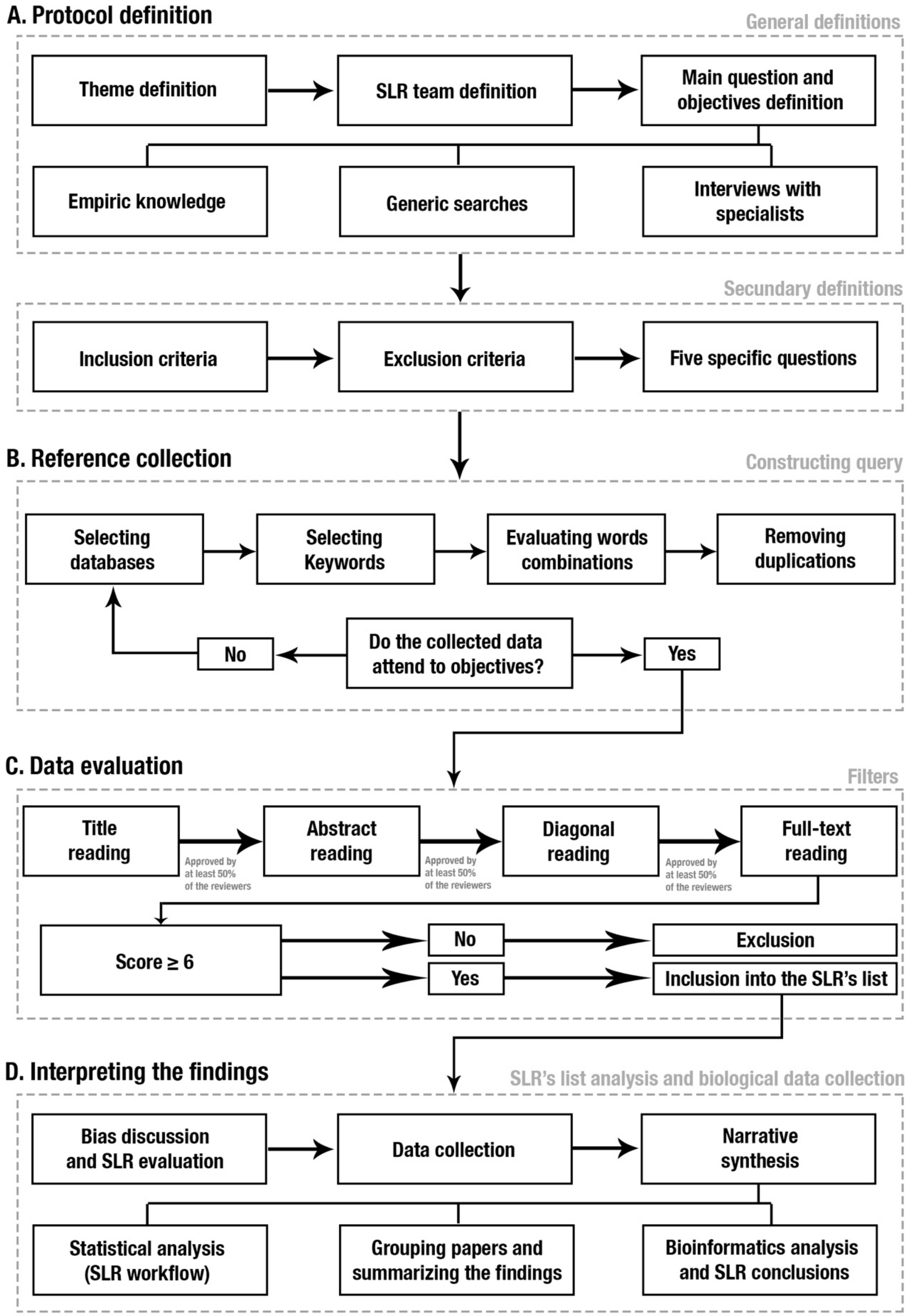

**Figure 2. BiSLR workflow.**

BiSLR is divided into four steps: (A) protocol definition; (B) reference collection; (C) data evaluation; and (D) interpreting the findings.

**Table 1. Protocol example.**

| | |
|---|---|
| **Main question** | "What is …" |
| **Objective** | "The purpose of this SLR is…" |
| **Inclusion criteria** | The criteria for inclusion are:<br>• Studies that mention […]<br>• Studies that contain the keywords: […] |
| **Exclusion criteria** | The criteria for exclusion are:<br>• Do not include the specific experiments […];<br>• Paper is not in English;<br>• Do not contain validation;<br>• Published data not available;<br>• Do not obtain 60% of the scoring in the last evaluation step (mandatory). |
| **Specific questions** | "The specific objectives are:"<br>• Question one: what is […]<br>• Question two: the research were performed in the period of […]<br>• Question three: the results contain experiments with […]<br>• Question four: there is *in vitro/in silico/in vivo* validation of […]?<br>• Question five: the database created in the study […] was available?" |

The main question will help to define the keywords used to search the databases. We advise the use of search strategies, such as search by complete expressions (in general, enclose the terms between quotation marks), searches for keywords in specific places of the manuscript (as title or abstract), and search manuscripts without undesired words. Different databases may present various methods for search. Hence, it is necessary to verify each database instruction manually to construct efficient queries.

Most databases offer downloadable textual, or table files of the search results. In addition, the collected references may be organized by software, such as Zotero (https://www.zotero.org) or Mendeley (https://www.mendeley.com/).

We point out that the process to perform an SLR must be iterative, *i.e.*, steps must be repeated until a final list of the best quality studies is accomplished. Hence, an SLR should be restarted at any moment that the protocol does not correspond to the objectives of the research. When an SRL is restarted, the data collected can be recycled. For instance, if some of the data for a SLR was already collected and analyzed, however after a few days a new and valuable reference is published. Following the rules, the SRL needs to be restarted, and all the steps redid. Nevertheless, the collected data can be recycled and the new data updated. Remember that an SLR must be rigorously registered, that even a person that did not participate in any step of the SLR could reproduce it.

### 2.3. Data evaluation

Data evaluation consists of the phase in which the reviewers analyze each reference collected and define the articles that will be included in the SLR list. We recommend that at least two researchers conduct the review independently. This process has been recommended to minimize the risk of selection bias in research [5].

BiSLR defines that data evaluation is performed in four substeps:

(i) **Title evaluation**: each reviewer evaluates if the article may be related to the main question based on the title. In this step, keywords related to the main question or objectives are searched in the title. The reviewer can approve or not the paper to be evaluated in the next step. If the evaluator has doubts about approving the study, it is recommended to accept the paper for a complete evaluation in the next step. This step is important to eliminate papers that do not have any relationship with the searched theme;

(ii) **Abstract evaluation**: the abstracts of the studied classified in the previous step are assessed in this step. In this step, it is important to evaluate if the paper complies with the main question. We recommended eliminating studies following the exclusion criteria. Once again, if the reviewer has doubts about classifying or not the paper for the next step, our suggestion is to approve it;

(iii) **Diagonal reading**: in this step, the reviewers must read the introduction, figures and tables titles, and conclusions of the papers classified for this step. In diagonal reading, the reviewers can improve their knowledge about the theme studied. Hence, in this step, the reviewers should verify if the specific questions attend to the objective of the SLR and describe better the main question. Also, they can verify if the keywords used in the database search were appropriated. If not, we recommend starting the SLR again, changing the keywords used in the search. This procedure could also be performed in the preceding steps, however, during the diagonal reading, the reviewers tend to have a higher knowledge about the theme. If any doubts arise about approving some of the studies, our instruction is to assembly all reviewers to discuss them. Doubts after the diagonal reading step may indicate that the SLR objective is not clear to all reviewers;

(iv) **Full-text reading**: in the final step of data evaluation, the reviewers must perform full-text reading and evaluate the studies by a scoring system. In this step, all reviewers must answer the five specific questions for each paper. For each specific question, a reviewer can give the scores: (2) if the study complies with the requirements of the question; (1) if the study partially satisfies the requirements of the question; or (0) if the study does not fulfill any of the question requirements. This evaluation is empirical, *i.e.* depends on the knowledge of the reviewer. If the paper has obtained a score equal or higher than six (60 %), it will be included in the SLR list.

A paper is accepted for the next SRL phase if it was approved by at least half of the reviewers. The exception is the full-text reading step, which also requires that the paper obtained a score higher than six.

### 2.4. Interpreting the finding

In the final phase of the SLR, the reviewers must interpret the finding. This phase is divided into six substeps: (i) group the references; (ii) summarizing the main topics; (iii) comparison among results; (iv) data collection; (v) evaluating the SLR; and (vi) narrative synthesis.

First, the reviewers have to group the papers included in the SLR list, summarize the main topics and compare the information obtained from different sources. This comparison may detect new finds, which could be the main results of an SLR. Also, a step of biological data collection can be performed at this stage. The papers may contain IDs of public biological databases, and the ID may be correlated to a function. The list of collected IDs can be used to construct curated databases for a specific topic. This outcome consists in the main impact for bioinformatics proposed by BiSLR.

In the end, BiSLR proposes that the SLR resulting be evaluated by the PRISMA method. This approach proposes that an SLR must exhibit a workflow summarizing numerically the papers collected by steps. In addition, the SLR quality should be evaluated by a list of topics required for an SLR. A PRISMA checklist may be constructed to assess the SLR. An example of PRISMA workflow and checklist is showed in the case study. We also recommend registering the SLR protocol. The register is necessary to guarantee that the procedures may be reproduced. Databases, such as PROSPERO (https://www.crd.york.ac.uk/PROSPERO/), allow the register of SLR protocols. Also, the SLR and database produced may be organized in a manuscript through the narrative synthesis phase and published as a review.

## 3. Case study

To evaluate our method, we proposed a case study to detect and summarize SLRs developed for bioinformatics. Four reviewers conducted the SLR case study, here called A, B, C and D. The case study shows a detailed step-by-step of how to proceed in a systematic literature review to collect data for bioinformatics analysis using the methods of BiSLR.

### 3.1. Protocol definition

The reviewers defined the protocol of the SLR case study (Table 2). For this case study, the goal was to search in the literature if examples of systematic literature reviews applied for bioinformatics analysis existed. The main question defined for the SLR was: "what is the state-of-the-art of the use of SLR in bioinformatics research?"

The specific questions defined were:

(i) Is the paper clearly described as an SLR?

- The description may appear in the title or the abstract.

(ii) Does the paper clearly present the restrictive methods of SLRs?

- Such as a list of papers collected and properly referenced; a quantitative evaluation of the collected data, for instance, PRISMA; a narrative synthesis of the results.

(iii)Does the paper main subject involve some bioinformatics study area?

- Such as genomics, transcriptomics, proteomics or metabolomics.

(iv) Is mentioned in the paper some OMICS database?

- Such as, GenBank [12], PDB [13,14], KEGG [15] or UniProt [16].

(v) Is cited in the paper any bioinformatics tool?

**Table 2. Protocol of the SLR case study.**

| | |
|---|---|
| **Main question** | What is the state-of-the-art of the use of SLR in bioinformatics research? |
| **Objective** | The objective was to search in the literature if examples of systematic literature reviews applied for bioinformatics analysis exist. |
| **Inclusion criteria** | The criteria for inclusion were:<br>• Studies that mention systematic analysis of the literature to collect biological data<br>• The paper describes some bioinformatics subarea, such as genomics, transcriptomics or proteomics;<br>• The paper suggests the use of bioinformatics tools to analyze biological data;<br>• The paper suggests the collection of data from biological databases;<br>• Studies that contain the keywords: bioinformatics and systematic literature review |
| **Exclusion criteria** | The criteria for exclusion were:<br>• The paper is not declared as a systematic literature review;<br>• Study is not in English;<br>• The paper does not describe any bioinformatics approach;<br>• Published data not available. |
| **Specific questions** | The specific questions were:<br>(i) Is the paper clearly described as an SLR?<br>(ii) Does the paper clearly present the restrictive methods of SLRs?<br>(iii) Does the paper main question involve any bioinformatics area?<br>(iv) Is mentioned in the paper some OMICS database?<br>(v) Is cited in the paper some bioinformatics tool? |

## 3.2. Reference collection

The keywords used in the query were defined in the reviewers meeting, and they were changed based on the first searches. We used two databases: PubMed (https://www.ncbi.nlm.nih.gov/pubmed/) and ScienceDirect (http://www.sciencedirect.com/). The searches were performed from July 2016 until August 2016. During this period, eight queries were tried (see Supplementary Material). The query for PubMed was: ("systematic review" [Title] OR "systematic literature review" [Title]) AND bioinformatics. For ScienceDirect the query used was: (TITLE("systematic literature review") OR TITLE("systematic review")) AND bioinformatics. A total of 232 references were collected from the databases (152 from PubMed and 80 from ScienceDirect). The records were downloaded in textual format. The references file was edited using Sublime Text software (one reference allocated in each line). A script developed in Python programming language was used to remove redundant references (see Supplementary Material). A total of 207 references were qualified for the data evaluation step.

### 3.3. Data evaluation

For each of the data evaluation four substeps, the reviewers gave a vote (yes or no) to approve or not the reference for the next phase. The data evaluation was described in the PRISMA workflow (Figure 3). In the title evaluation, the reviewers A, B, C and D approved 55, 72, 24 and 34 papers, respectively. A total of 59 references were approved by at least two evaluators (159 were eliminated). In the abstract evaluation, the reviewers approved 32, 34, 25 and 32 papers, respectively. A total of 33 references were approved for the next step (26 were eliminated). In the diagonal reading substep, the reviewers approved 10, 12, nine and nine papers, respectively (12 classified and 21 eliminated). In the full-text reading, the reviewers approved nine, nine, eight and eight papers. From nine papers approved by at least two reviewers, one study did not obtain a score equal or higher than six (the complete matrix is available in the Supplementary Material).

### 3.4. Interpreting the finding

The papers were grouped based on bioinformatics sub-area: genomics, transcriptomics, and proteomics. In addition, the reviewers compared the bioinformatics tools presented in the papers and collected IDs from biological databases.

### 3.5. Narrative synthesis

Eight papers were included in the SLR list (Table 3). The papers covered several areas, such as heart diseases [17], bacterial infections [18–20], tendinopathy [21] and cancer [22–24]. We grouped the papers based on the type of OMICS: genomics [18–20], transcriptomics [17,22,24], and proteomics [21,23] (Table 4). However, some papers mentioned more than one type of OMICS, such as transcriptomics and proteomics [21], genomics and transcriptomics [23] and a combination of the three types [17,22].

Overall, six of the eight selected studies mentioned or made use of an OMICS database (Table 4). Since the remaining studies dealt with various subjects, no consensus between databases was evident. However, Genbank was referred to in three studies, while UniGene in two. Also, it is unclear which specific EMBL and NCBI databases were used in Sharma et al. [17] and Liu et al. [23], respectively, since both account for numerous databases. Bioinformatics tools were mentioned or used in five of the eight studies (Table 4). A description of the software and web servers can be found in the Supplementary Material. None of the tools appeared in more than one study. Again, this might be a consequence of the diverse subjects broached in the studies. Interestingly, the preferred databases in Nikolayevskyy et al. [18], MIRU-VNTRplus and SITVIT-WEB, contain online tools to analyze the chosen data, making them a bioinformatics application as well. In the end, it seemed clear that a wide number of tools can be employed and they may play important and different roles in OMICS research.

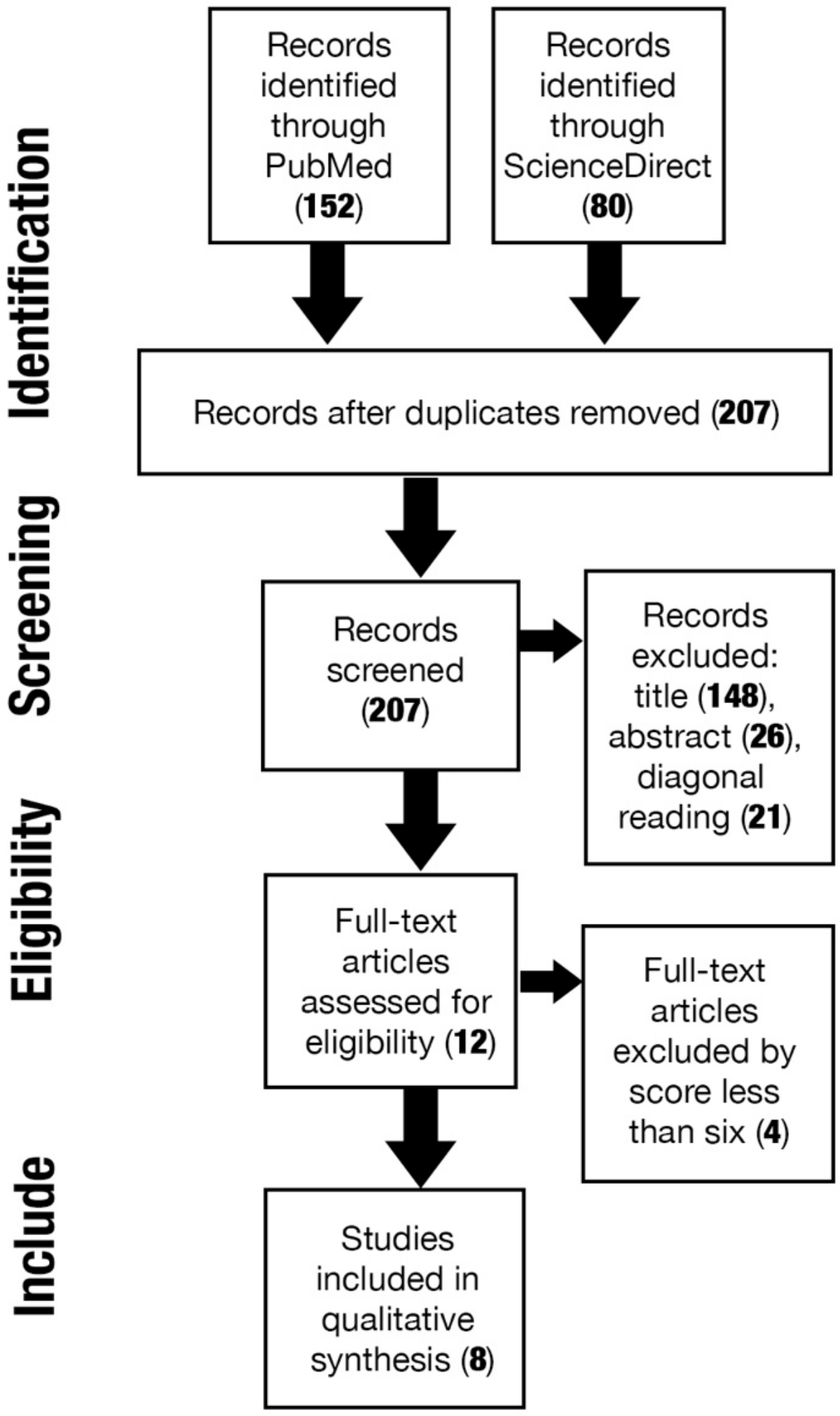

**Figure 3. PRISMA workflow. Describing the number of papers collected during the systematic literature review according to PRISMA statement.**

### 3.5. Discussion

The SLR case study was used to demonstrate how to proceed in systematic literature reviews. It was preceded by four reviewers, which reduced the bias risk. The data collected was heterogeneous. Therefore, it is not useful to construct a database and perform bioinformatics analysis with its outcome. Also, the keywords used were generic and related to the theme, and this may be a bias risk for the SLR case study.

The studies collected could be classified into three main groups: genomics, transcriptomics, and proteomics. Indeed, the bioinformatics is strongly related to OMICS sciences. Although the word bioinformatics had several meanings throughout the decades, nowadays it might be possible to suggest that bioinformatics is the use of computational tools to threat OMICS data.

**Table 3. List of papers inserted in the SLR case study.**

| # | Paper | Score | Source |
|---|---|---|---|
| **1** | A systematic review of large scale and heterogeneous gene array data in heart failure | 9.75 | [17] |
| **2** | Whole genome sequencing of Mycobacterium tuberculosis for detection of recent transmission and tracing outbreaks: A systematic review | 7.75 | [18] |
| **3** | Systematic review of mutations in pyrazinamidase associated with pyrazinamide resistance in Mycobacterium tuberculosis clinical isolates | 7.5 | [19] |
| **4** | Proteomics perspectives in rotator cuff research: a systematic review of gene expression and protein composition in human tendinopathy | 6 | [21] |
| **5** | Genetic susceptibility to invasive meningococcal disease: MBL2 structural polymorphisms revisited in a large case-control study and a systematic review | 9 | [20] |
| **6** | Melanoma prognosis: a REMARK-based systematic review and bioinformatic analysis of immunohistochemical and gene microarray studies. | 8 | [22] |
| **7** | Upregulated and downregulated proteins in hepatocellular carcinoma: a systematic review of proteomic profiling studies | 9 | [23] |
| **8** | The evolving transcriptome of head and neck squamous cell carcinoma: a systematic review | 9.75 | [24] |

**Table 4. Databases and bioinformatics tools mentioned or used in the selected studies.**

| # | OMICS | Database | Bioinformatics tools | Reference |
|---|---|---|---|---|
| **1** | Transcriptomics | dbEst;<br>EMBL databases;<br>UniGene;<br>GenBank. | GenMAPP (Gene MicroArray Pathway Profiler);<br>MAPPFinder;<br>GoMiner;<br>FatiGo;<br>OntoExpress;<br>GoSurfer;<br>SignalP 1.1 Server. | [17] |
| **2** | Genomics | MIRU-VNTRplus;<br>SITVIT-WEB. | MIRU-VNTRplus;<br>SITVIT-WEB. | [18] |
| **3** | Genomics | GenBank. | - | [19] |
| **4** | Proteomics | - | - | [21] |
| **5** | Genomics | WTCCC (Welcome Trust Case-Control Consortium). | SNPforID browser;<br>GeneMarker;<br>STRUCTURE v2.3.3;<br>PLINK v1.07;<br>EIGENSOFT v2.0;<br>SEQUENCHER v4.9;<br>EPI INFO v6.0. | [20] |
| **6** | Transcriptomics | - | MetaCore. | [22] |
| **7** | Proteomics | Swiss-Prot;<br>NCBI databases;<br>UniProt-GOA;<br>IPI database. | - | [23] |
| **8** | Transcriptomics | UniGene;<br>Genbank. | Cytoscape;<br>Ingenuity Pathways Analysis. | [24] |

## 4. Conclusions

We detected few SLR studies for bioinformatics. Only eight papers fulfilled the required criteria and included in the SLR. To our best knowledge, BiSLR is the first report of an SLR method for bioinformatics studies. We expect that BiSLR will allow the increase of SLR papers for bioinformatics analysis.

## Acknowledgements

The authors thank the funding agencies: Coordenação de Aperfeiçoamento de Pessoal de Nível Superior (CAPES), Fundação de Amparo a Pesquisa do Estado de Minas Gerais (FAPEMIG) and Conselho Nacional de Desenvolvimento Científico e Tecnológico (CNPq). Authors would like to thank Larissa Leijôto. This study was funded by Coordenação de Aperfeiçoamento de Pessoal de Nível Superior (CAPES). Project number: 51/2013 - 23038.004007/2014-82.

# Supplementary material

## Script 1. Script to remove repeated references.

Script to remove repeated references in the list of collected references. The script, developed using Python, analyzes repetitive DOIs numbers. It receives a textual file where each citation needs to be in a line. It returns a file, called “unique_references.txt”, without redundant citations. It also shows possible redundant lines that the script cannot analyze with precision. To run, execute in a terminal or python CLI: “python removeDuplications.py”. The script requires a file called “references.txt” in the same directory (where each reference need to be in one line).

```
# removeDuplications.py
# Function: receive a list of references (each one in a line) and remove duplications
# Author: Diego Mariano
# Date: August 19, 2016

# Imports
import re
import sys

print "*** Remove Duplications ***"

# Files
files = sys.argv
if len(files) == 1:
	files = [‘references.txt’]


# Variables
DOIs = []
unique = []
warnings = []
eliminated = []


# For each file
for i in range(len(files)):

	# open file
	lines = open(files[i]).readlines()

	# Add all first file
	for j in range(len(lines)):

		# Remove special characters
		lines[j].replace("\r"," ")
		lines[j].replace("\t"," ")

		# Collect DOI
		doi_obj = re.search('[0-9][0-9]\.[0-9][0-9][0-9][0-9]([0-9]?)\/([a-z]|[A-Z]|\.|[0-9]|\(|\)|\-
|\_|\/)*([0-9]|[a-z])(\. | )', lines[j])
```

```
		try:
			doi = doi_obj.group(0)
		except:
			doi = ""
			warnings.append(j+1)

		# Remove spaces
		doi = doi.strip()

		# Remove dot
		if doi[-1:] == ".":
			doi = doi[:-1]

		if doi != "" and doi not in DOIs:
			DOIs.append(doi)
			unique.append(lines[j])
		elif doi == "":
			unique.append(lines[j])
		elif doi in DOIs:
			eliminated.append(j+1)

print "\n"+str(len(DOIs))+" unique DOIs were detected."

print "\n"+str(len(warnings))+" warnings were detected on the lines: "
print warnings
print "Please, verify manually."

print "\n"+str(len(eliminated))+" duplications removed were detected on the lines: "
print eliminated

print "\nCreating the file: unique_references.txt"

w = open("unique_references.txt","w")
for line in unique:
	w.write(line)

w.close()

print "\nSuccessful"
```

## Supplementary Table S1. List of databases.

| Database | Target public | Link |
|---|---|---|
| *PubMed* | PubMed includes citations for biomedical literature from MEDLINE, life science journals, and online books. | http://www.ncbi.nlm.nih.gov/pubmed |
| *ScienceDirect* | Database of scientific, technical, and medical researches | http://www.sciencedirect.com/ |
| *Scopus* | Scopus is described as the largest abstract and citation database of peer-reviewed literature: books, conference proceedings, and scientific journals. | https://www.scopus.com/ |
| *Web of Science* | Web of Science provides access to the most reliable, integrated, multidisciplinary research connected through linked content citation metrics from multiple sources within a single interface. | https://webofknowledge.com |
| *Journal Citation Reports (JCR)* | JCR helps to measure research influence and impact at the journal and category levels and shows the relationship between citing and cited journals. | https://jcr.incites.thomsonreuters.com |
| *Engineering Village* | Engineering Village, the essential engineering research database, provides a searchable index of the most comprehensive engineering literature and patent information available. | https://www.engineeringvillage.com |
| *MathSciNet* | MathSciNet® is an electronic publication offering access to a carefully maintained and easily searchable database of reviews, abstracts and bibliographic information for much of the mathematical sciences literature. | http://www.ams.org/mathscinet/ |
| *Begell House Digital Library* | The Begell Digital Library Journals Collection is a consolidated and peer-reviewed collection of leading academic, research, and applied work in the fields of engineering and biomedical sciences. | http://www.dl.begellhouse.com/ |
| *IEEE Xplore Digital Library* | IEEE *Xplore* provides web access to more than three-million full-text documents from some of the world's most highly cited publications in electrical engineering, computer science, and electronics. | http://ieeexplore.ieee.org/ |
| *ACM Digital Library* | The ACM Digital Library is a research, discovery and networking platform containing: the Full-Text Collection of all ACM publications, including journals, conference proceedings, technical magazines, newsletters and books; and a collection of curated and hosted full-text publications from select publishers for Computing Literature. | http://dl.acm.org/ |
| *Springer* | A database with more than 2,900 journals and 200,000 books. | http://www.springer.com/ |
| CrossRef | Search Crossref's database of 80 million records for authors, titles, DOIs, ORCIDs, ISSNs, funders, license URIs, etc. | http://www.crossref.org/ |

## Supplementary Table S2. Iterative queries performed in the databases.

The query for "systematic" and "review" separately returns undesirable results, such as the expression "review on systematic risk assessment". In some cases, it was necessary to use different queries for the databases. In this case, the query was descript as PM (PubMed) or SD (ScienceDirect). Search performed on July 1, 2016.

| | | Query | PubMed | ScienceDirect | Total |
|---|---|---|---|---|---|
| **1** | systematic literature review AND bioinformatics | | 2,719 | 2,880 | 5,599 |
| **2** | "systematic literature review" AND bioinformatics | | 12 | 80 | 92 |
| **3** | ("systematic literature review" OR SLR) AND bioinformatics | | 28 | 184 | 212 |
| **4** | ("systematic literature review" OR "systematic review") AND bioinformatics | | 219 | 807 | 1,026 |
| **5** | **PM** | (systematic[Title] AND review[Title]) AND bioinformatics | 156 | 80 | 236 |
| | **SD** | TITLE(**systematic**) AND TITLE(**review**) AND bioinformatics | | | |
| **6** | **PM** | ("systematic review" [Title]) AND bioinformatics | 146 | 72 | 218 |
| | **SD** | TITLE("systematic review") AND bioinformatics | | | |
| **7** | **PM** | ("systematic literature review" [Title]) AND bioinformatics | 6 | 8 | 14 |
| | **SD** | TITLE("systematic literature review") AND bioinformatics | | | |
| **8** | **PM** | ("systematic review" [Title] OR "systematic literature review" [Title]) AND bioinformatics | 152 | 80 | 232 |
| | **SD** | (TITLE("systematic literature review") OR TITLE("systematic review")) AND bioinformatics | | | |

## Supplementary Table S3. Bioinformatics tools cited in the SLRs collected.

| **Bioinformatics tools** | **Description** |
|---|---|
| **GenMAPP** | A program for viewing and analyzing microarray data on microarray pathway profiles (MAPPs) representing biological pathways or any other functional grouping of genes. |
| **MAPPFinder** | A tool that dynamically links gene expression data to the gene ontogeny (GO) hierarchy. |
| **GoMiner** | A tool to interpret genomic as well as proteomic data in the context of gene ontology. |
| **FatiGo** | A web tool for finding significant associations of Gene Ontology terms with groups of genes. |
| **OntoExpress** | A tool able to automatically translate lists of differentially regulated genes into functional profiles characterizing the impact of the condition studied. |
| **GoSurfer** | A graphical interactive tool for comparative analysis of large gene sets in Gene Ontology space. |
| **SignalP Server** | A web server predicts the presence and location of signal peptide cleavage sites in amino acid sequences from different organisms. |
| **MIRU-VNTRplus** | Reference database and analysis tool for Mycobacterium tuberculosis *MIRU*-VNTR (MLVA)-, Spoligo-, RD (LSP)-, and SNP-data. |
| **SITVIT-WEB** | SITVIT is a **Mycobacterium tuberculosis** molecular markers database. Query and analysis tools can be used to access the contents of this database using different criteria. |
| **SNPforID browser** | A web framework for any given population based SNP genotype database that can deal even with virtually any number of genotypes, and that is also capable of summarizing all that information into the most common population genetics indices. |
| **GeneMarker** | A unique genotype analysis software which integrates new technologies enhancing the speed, accuracy, and ease of analysis. |
| **STRUCTURE** | The program implements a model-based clustering method for inferring population structure using genotype data consisting of unlinked markers. |
| **PLINK** | A free, open-source whole genome association analysis toolset, designed to perform a range of basic, large-scale analyses in a computationally efficient manner. |
| **EIGENSOFT** | A package that combines functionality from population genetics methods and stratification correction method. The method uses principal components analysis to explicitly model ancestry differences between cases and controls along continuous axes of variation. |
| **SEQUENCHER** | A DNA sequencing tool written where users enter or import sequence fragments. Sequences can then be screened for vector contamination using Sequencher's sensitive vector screening. |
| **EPI INFO** | A public domain suite of interoperable software tools designed for the global community of public health practitioners and researchers. It provides for easy data entry form and database construction, a customized data entry experience, and data analyses with epidemiologic statistics, maps, and graphs for public health professionals who may lack an information technology background. |
| **MetaCore** | High-quality biological systems content in context, giving essential data and analytical tools to accelerate scientific research. |
| **Cytoscape** | Cytoscape is an open source software platform for visualizing molecular interaction networks and biological pathways and integrating these networks with annotations, gene expression profiles, and other state data. |
| **Ingenuity Pathways Analysis** | A powerful analysis and search tool that uncovers the significance of omics data and identifies new targets or candidate biomarkers within the context of biological systems. |

**Supplementary Table S4. PRIMA checklist.**

| Section/topic | # | Checklist item | Reported on page # |
|---|---|---|---|
| **TITLE** | | | |
| Title | 1 | Identify the report as a systematic review, meta-analysis, or both. | 1 |
| **ABSTRACT** | | | |
| Structured summary | 2 | Provide a structured summary including, as applicable: background; objectives; data sources; study eligibility criteria, participants, and interventions; study appraisal and synthesis methods; results; limitations; conclusions and implications of key findings; systematic review registration number. | 3 |
| **INTRODUCTION** | | | |
| Rationale | 3 | Describe the rationale for the review in the context of what is already known. | 4,5 |
| Objectives | 4 | Provide an explicit statement of questions being addressed with reference to participants, interventions, comparisons, outcomes, and study design (PICOS). | 11 |
| **METHODS** | | | |
| Protocol and registration | 5 | Indicate if a review protocol exists, if and where it can be accessed (e.g., Web address), and, if available, provide registration information including registration number. | 11,12,21 |
| Eligibility criteria | 6 | Specify study characteristics (e.g., PICOS, length of follow-up) and report characteristics (e.g., years considered, language, publication status) used as criteria for eligibility, giving rationale. | 12,21 |
| Information sources | 7 | Describe all information sources (e.g., databases with dates of coverage, contact with study authors to identify additional studies) in the search and date last searched. | 12,21 |
| Search | 8 | Present full electronic search strategy for at least one database, including any limits used, such that it could be repeated. | 12 |
| Study selection | 9 | State the process for selecting studies (i.e., screening, eligibility, included in systematic review, and, if applicable, included in the meta-analysis). | 12,13 |
| Data collection process | 10 | Describe method of data extraction from reports (e.g., piloted forms, independently, in duplicate) and any processes for obtaining and confirming data from investigators. | 12,13 |
| Data items | 11 | List and define all variables for which data were sought (e.g., PICOS, funding sources) and any assumptions and simplifications made. | 21 |
| Risk of bias in individual studies | 12 | Describe methods used for assessing risk of bias of individual studies (including specification of whether this was done at the study or outcome level), and how this information is to be used in any data synthesis. | 11 |
| Summary measures | 13 | State the principal summary measures (e.g., risk ratio, difference in means). | 13,22 |
| Synthesis of results | 14 | Describe the methods of handling data and combining results of studies, if done, including measures of consistency (e.g., $I^2$) for each meta-analysis. | 13,22 |
| Risk of bias across studies | 15 | Specify any assessment of risk of bias that may affect the cumulative evidence (e.g., publication bias, selective reporting within studies). | - |
| Additional analyses | 16 | Describe methods of additional analyses (e.g., sensitivity or subgroup analyses, meta-regression), if done, indicating which were pre-specified. | - |
| **RESULTS** | | | |
| Study selection | 17 | Give numbers of studies screened, assessed for eligibility, and included in the review, with reasons for exclusions at each stage, ideally with a flow diagram. | 19 |
| Study characteristics | 18 | For each study, present characteristics for which data were extracted (e.g., study size, PICOS, follow-up period) and provide the citations. | 13,14,23 |
| Risk of bias within studies | 19 | Present data on risk of bias of each study and, if available, any outcome- level assessment (see item 12). | 22 |
| Results of individual studies | 20 | For all outcomes considered (benefits or harms), present, for each study: (a) simple summary data for each intervention group (b) effect estimates and confidence intervals, ideally with a forest plot. | 13,14,23 |
| Synthesis of results | 21 | Present results of each meta-analysis done, including confidence intervals and measures of consistency. | 13,14,23 |
| Risk of bias across studies | 22 | Present results of any assessment of risk of bias across studies (see Item 15). | - |
| Additional analysis | 23 | Give results of additional analyses, if done (e.g., sensitivity or subgroup analyses, meta-regression [see Item 16]). | - |
| **DISCUSSION** | | | |
| Summary of evidence | 24 | Summarize the main findings including the strength of evidence for each main outcome; consider their relevance to key groups (e.g., healthcare providers, users, and policy makers). | 14 |
| Limitations | 25 | Discuss limitations at study and outcome level (e.g., risk of bias), and at review-level (e.g., incomplete retrieval of identified research, reporting bias). | 14 |
| Conclusions | 26 | Provide a general interpretation of the results in the context of other evidence, and implications for future research. | 14,15 |
| **FUNDING** | | | |
| Funding | 27 | Describe sources of funding for the systematic review and other support (e.g., supply of data); role of funders for the systematic review. | 15 |

**Supplementary Table S5. Empirical score evaluation of references classified for the step four.**

References of #1-2 were obtained from ScienceDirect; references of #3-12 were obtained from PubMed. Four evaluators (A, B, C, and D) analyzed the 12 references. They answered five questions for each paper and gave three possible scores: (0) the paper does not comply with the requirements of the question; (1) the paper attends complies with the requirements of the question partially; or (2) the paper fulfills the requirements of the question. The score evaluation was empirical, *i.e.* it depends on the evaluator's opinion. Eight papers obtained a score higher than sixty percent, and they were included in the SLR.

| # | References | Evaluators | | | | | | | | | | | | | | | | | | | | | | | | |
|---|---|---|---|---|---|---|---|---|---|---|---|---|---|---|---|---|---|---|---|---|---|---|---|---|---|---|
| | | A | | | | | | B | | | | | | C | | | | | | D | | | | | | Score average |
| | Questions: | 1 | 2 | 3 | 4 | 5 | # | 1 | 2 | 3 | 4 | 5 | # | 1 | 2 | 3 | 4 | 5 | # | 1 | 2 | 3 | 4 | 5 | # | |
| **1** | Umesh C. Sharma, Saraswati Pokharel, Chris T.A. Evelo, Jos G. Maessen, A systematic review of large scale and heterogeneous gene array data in heart failure, Journal of Molecular and Cellular Cardiology, Volume 38, Issue 3, March 2005, Pages 425-432, ISSN 0022-2828, http://dx.doi.org/10.1016/j.yjmcc.2004.12.016. (http://www.sciencedirect.com/science/article/pii/S0022282804004109) | 2 | 2 | 2 | 2 | 2 | **10** | 2 | 2 | 2 | 2 | 2 | **10** | 2 | 1 | 2 | 2 | 2 | **9** | 2 | 2 | 2 | 2 | 2 | **10** | **9.75** |
| **2** | Vlad Nikolayevskyy, Katharina Kranzer, Stefan Niemann, Francis Drobniewski, Whole genome sequencing of Mycobacterium tuberculosis for detection of recent transmission and tracing outbreaks: A systematic review, Tuberculosis, Volume 98, May 2016, Pages 77-85, ISSN 1472-9792, http://dx.doi.org/10.1016/j.tube.2016.02.009. (http://www.sciencedirect.com/science/article/pii/S1472979215302705) | 2 | 2 | 2 | 1 | 2 | **9** | 2 | 2 | 2 | 1 | 0 | **7** | 2 | 2 | 2 | 0 | 2 | **8** | 2 | 2 | 2 | 0 | 1 | **7** | **7.75** |
| **3** | Ramirez-Busby SM, Valafar F. Systematic review of mutations in pyrazinamidase associated with pyrazinamide resistance in Mycobacterium tuberculosis clinical isolates. Antimicrob Agents Chemother. 2015 Sep;59(9):5267-77. doi: 10.1128/AAC.00204-15. Epub 2015 Jun 15. Review. PubMed PMID: 26077261; PubMed Central PMCID: PMC4538510. | 2 | 2 | 2 | 2 | 0 | **8** | 2 | 2 | 1 | 1 | 0 | **6** | 2 | 1 | 2 | 2 | 0 | **7** | 2 | 2 | 2 | 2 | 1 | **9** | **7.5** |
| **4** | Sejersen MH, Frost P, Hansen TB, Deutch SR, Svendsen SW. Proteomics perspectives in rotator cuff research: a systematic review of gene expression and protein composition in human tendinopathy. PLoS One. 2015 Apr 16;10(4):e0119974. doi: 10.1371/journal.pone.0119974. eCollection 2015. Review. PubMed PMID: 25879758; PubMed Central PMCID: PMC4400011. | 2 | 2 | 2 | 0 | 0 | **6** | 2 | 2 | 2 | 0 | 1 | **7** | 2 | 2 | 2 | 0 | 0 | **6** | 2 | 2 | 1 | 0 | 0 | **5** | **6** |
| **5** | Kaufman JS, Dolman L, Rushani D, Cooper RS. The contribution of genomic research to explaining racial disparities in cardiovascular disease: a systematic review. Am J Epidemiol. 2015 Apr 1;181(7):464-72. doi: 10.1093/aje/kwu319. Epub 2015 Mar 1. Review. PubMed PMID: | 2 | 2 | 2 | 0 | 0 | **6** | 2 | 1 | 2 | 2 | 0 | **7** | 2 | 1 | 1 | 0 | 0 | **4** | 2 | 2 | 1 | 0 | 0 | **5** | **5.5** |
| | 25731887. | | | | | | | | | | | | | | | | | | | | | | | | | |
| **6** | O'Mara-Eves A, Thomas J, McNaught J, Miwa M, Ananiadou S. Using text mining for study identification in systematic reviews: a systematic review of current approaches. Syst Rev. 2015 Jan 14;4:5. doi: 10.1186/2046-4053-4-5. Review. Erratum in: Syst Rev. 2015;4:59. PubMed PMID: 25588314; PubMed Central PMCID: PMC4320539. | 2 | 2 | 0 | 0 | 1 | **5** | 2 | 1 | 0 | 0 | 0 | **3** | 2 | 1 | 0 | 0 | 0 | **3** | 2 | 2 | 0 | 0 | 0 | **4** | **3.75** |
| **7** | Shabani M, Bezuidenhout L, Borry P. Attitudes of research participants and the general public towards genomic data sharing: a systematic literature review. Expert Rev Mol Diagn. 2014 Nov;14(8):1053-65. doi: 10.1586/14737159.2014.961917. Epub 2014 Sep 26. Review. PubMed PMID: 25260013. | 2 | 2 | 1 | 0 | 0 | **5** | 2 | 2 | 1 | 0 | 0 | **5** | 2 | 2 | 1 | 0 | 0 | **5** | 2 | 2 | 1 | 1 | 0 | **6** | **5.25** |
| **8** | Wang X, Zhang A, Sun H, Wang P. Systems biology technologies enable personalized traditional Chinese medicine: a systematic review. Am J Chin Med. 2012;40(6):1109-22. doi: 10.1142/S0192415X12500826. Review. PubMed PMID: 23227785. | 2 | 0 | 2 | 0 | 0 | **4** | 2 | 0 | 2 | 0 | 0 | **4** | 2 | 0 | 2 | 0 | 0 | **4** | 2 | 0 | 1 | 0 | 0 | **3** | **3.75** |
| **9** | Bradley DT, Bourke TW, Fairley DJ, Borrow R, Shields MD, Young IS, Zipfel PF, Hughes AE. Genetic susceptibility to invasive meningococcal disease: MBL2 structural polymorphisms revisited in a large case-control study and a systematic review. Int J Immunogenet. 2012 Aug;39(4):328-37. doi: 10.1111/j.1744-313X.2012.01095.x. Epub 2012 Feb 2. Review. PubMed PMID: 22296677. | 2 | 2 | 2 | 2 | 2 | **10** | 2 | 2 | 2 | 2 | 2 | **10** | 2 | 2 | 2 | 1 | 1 | **8** | 2 | 2 | 2 | 1 | 1 | **8** | **9** |
| **10** | Schramm SJ, Mann GJ. Melanoma prognosis: a REMARK-based systematic review and bioinformatic analysis of immunohistochemical and gene microarray studies. Mol Cancer Ther. 2011 Aug;10(8):1520-8. doi: 10.1158/1535-7163.MCT-10-0901. Epub 2011 Jun 9. Review. PubMed PMID: 21659462. | 2 | 1 | 2 | 1 | 2 | **8** | 2 | 1 | 2 | 1 | 2 | **8** | 2 | 1 | 2 | 0 | 2 | **7** | 2 | 2 | 2 | 1 | 2 | **9** | **8** |
| **11** | Liu Z, Ma Y, Yang J, Qin H. Upregulated and downregulated proteins in hepatocellular carcinoma: a systematic review of proteomic profiling studies. OMICS. 2011 Jan-Feb;15(1-2):61-71. doi: 10.1089/omi.2010.0061. Epub 2010 Aug 20. Review. PubMed PMID: 20726783. | 2 | 2 | 2 | 2 | 2 | **10** | 2 | 2 | 2 | 1 | 1 | **8** | 2 | 2 | 2 | 1 | 1 | **8** | 2 | 2 | 2 | 2 | 2 | **10** | **9** |
| **12** | Yu YH, Kuo HK, Chang KW. The evolving transcriptome of head and neck squamous cell carcinoma: a systematic review. PLoS One. 2008 Sep 15;3(9):e3215. doi: 10.1371/journal.pone.0003215. PubMed PMID: 18791647; PubMed Central PMCID: PMC2533097. | 2 | 2 | 2 | 2 | 2 | **10** | 2 | 2 | 2 | 2 | 2 | **10** | 2 | 2 | 2 | 2 | 1 | **9** | 2 | 2 | 2 | 2 | 2 | **10** | **9.75** |

## Supplementary Table S6. Evaluation of 207 references collected in the SLR.

| # | Reference | Step 1 | | | | | Step 2 | | | | | Step 3 | | | | | Step 4 | | | | |
|---|---|---|---|---|---|---|---|---|---|---|---|---|---|---|---|---|---|---|---|---|---|
| | | A | B | C | D | # | A | B | C | D | # | A | B | C | D | # | A | B | C | D | # |
| **1** | Umesh C. Sharma, Saraswati Pokharel, Chris T.A. Evelo, Jos G. Maessen, A systematic review of large scale and heterogeneous gene array data in heart failure, Journal of Molecular and Cellular Cardiology, Volume 38, Issue 3, March 2005, Pages 425-432, ISSN 0022-2828, http://dx.doi.org/10.1016/j.yjmcc.2004.12.016. (http://www.sciencedirect.com/science/article/pii/S0022282804004109) Keywords: Microarray; Heart failure; Bioinformatics; Expressed sequence tags | no | yes | yes | yes | 3 | yes | yes | yes | yes | 4 | yes | yes | yes | yes | 4 | yes | yes | yes | yes | 4 |
| **2** | Roger Colobran, Mireia Giménez-Barcons, Ana Marín-Sánchez, Eduard Porta-Pardo, Ricardo Pujol-Borrell, AIRE genetic variants and predisposition to polygenic autoimmune disease: The case of Graves' disease and a systematic literature review, Human Immunology, Available online 4 June 2016, ISSN 0198-8859, http://dx.doi.org/10.1016/j.humimm.2016.06.002. (http://www.sciencedirect.com/science/article/pii/S0198885916301112) Keywords: AIRE; Graves' disease; Autoimmunity; Mutation; Polymorphism; APECED | no | yes | no | yes | 2 | yes | yes | no | yes | 3 | yes | no | no | no | 1 | no | no | no | no | 0 |
| **3** | Dustin Heaton, Jeffrey C. Carver, Claims about the use of software engineering practices in science: A systematic literature review, Information and Software Technology, Volume 67, November 2015, Pages 207-219, ISSN 0950-5849, http://dx.doi.org/10.1016/j.infsof.2015.07.011. (http://www.sciencedirect.com/science/article/pii/S0950584915001342) Keywords: Computational science; Systematic literature review; Scientific software | yes | no | yes | no | 2 | no | no | no | no | 0 | no | no | no | no | 0 | no | no | no | no | 0 |
| **4** | Zohreh Sharafi, Zéphyrin Soh, Yann-Gaël Guéhéneuc, A systematic literature review on the usage of eye-tracking in software engineering, Information and Software Technology, Volume 67, November 2015, Pages 79-107, ISSN 0950-5849, http://dx.doi.org/10.1016/j.infsof.2015.06.008. (http://www.sciencedirect.com/science/article/pii/S0950584915001196) Keywords: Eye-tracking; Software engineering; Experiment | no | no | no | no | 0 | no | no | no | no | 0 | no | no | no | no | 0 | no | no | no | no | 0 |
| **5** | Upulee Kanewala, James M. Bieman, Testing scientific software: A systematic literature review, Information and Software Technology, Volume 56, Issue 10, | yes | no | yes | no | 2 | no | no | no | yes | 1 | no | no | no | no | 0 | no | no | no | no | 0 |
| | October 2014, Pages 1219-1232, ISSN 0950-5849, http://dx.doi.org/10.1016/j.infsof.2014.05.006. (http://www.sciencedirect.com/science/article/pii/S0950584914001232) Keywords: Scientific software; Software testing; Systematic literature review; Software quality | | | | | | | | | | | | | | | | | | | | |
| **6** | Ivanilton Polato, Reginaldo Ré, Alfredo Goldman, Fabio Kon, A comprehensive view of Hadoop research—A systematic literature review, Journal of Network and Computer Applications, Volume 46, November 2014, Pages 1-25, ISSN 1084-8045, http://dx.doi.org/10.1016/j.jnca.2014.07.022. (http://www.sciencedirect.com/science/article/pii/S1084804514001635) Keywords: Systematic literature review; Apache Hadoop; MapReduce; HDFS; Survey | no | no | no | no | 0 | no | no | no | no | 0 | no | no | no | no | 0 | no | no | no | no | 0 |
| **7** | Clara D.M. van Karnebeek, Sylvia Stockler, Treatable inborn errors of metabolism causing intellectual disability: A systematic literature review, Molecular Genetics and Metabolism, Volume 105, Issue 3, March 2012, Pages 368-381, ISSN 1096-7192, http://dx.doi.org/10.1016/j.ymgme.2011.11.191. (http://www.sciencedirect.com/science/article/pii/S1096719211006081) Keywords: Inborn errors of metabolism; Intellectual disability; Developmental delay; Therapy; Evidence; Systematic review | no | no | no | no | 0 | no | no | no | no | 0 | no | no | no | no | 0 | no | no | no | no | 0 |
| **8** | George Pentheroudakis, F.A. Greco, Nicholas Pavlidis, Molecular assignment of tissue of origin in cancer of unknown primary may not predict response to therapy or outcome: A systematic literature review, Cancer Treatment Reviews, Volume 35, Issue 3, May 2009, Pages 221-227, ISSN 0305-7372, http://dx.doi.org/10.1016/j.ctrv.2008.10.003. (http://www.sciencedirect.com/science/article/pii/S030573720800296X) Keywords: Cancer of unknown primary; Molecular profiling; Chemotherapy; Response; Survival | no | yes | no | no | 1 | no | no | no | no | 0 | no | no | no | no | 0 | no | no | no | no | 0 |
| **9** | Muhammad Ali Khan, Omair Atiq, Nisa Kubiliun, Bilal Ali, Faisal Kamal, Richard Nollan, Mohammad Kashif Ismail, Claudio Tombazzi, Michel Kahaleh, Todd H. Baron, Efficacy and safety of endoscopic gallbladder drainage in acute cholecystitis: Is it better than percutaneous gallbladder drainage? A systematic review and meta-analysis, Gastrointestinal Endoscopy, Available online 22 June 2016, ISSN 0016-5107, http://dx.doi.org/10.1016/j.gie.2016.06.032. (http://www.sciencedirect.com/science/article/pii/S0016510716302784) | no | no | no | no | 0 | no | no | no | no | 0 | no | no | no | no | 0 | no | no | no | no | 0 |
| **10** | Anna Jo Smith, Elizabeth L. Turner, Sanjay Kinra, Universal Cholesterol Screening in Childhood: A Systematic Review, Academic Pediatrics, Available online 21 June 2016, ISSN 1876-2859, http://dx.doi.org/10.1016/j.acap.2016.06.005. (http://www.sciencedirect.com/science/article/pii/S187628591630331X) Mesh Key Words: WordsHypercholesterolemia; Dyslipidemias; Cardiovascular diseases; Mass screening; Pediatrics; Primary healthcare; Patient Acceptance of Healthcare; Healthcare costs; Health services research | no | no | no | no | 0 | no | no | no | no | 0 | no | no | no | no | 0 | no | no | no | no | 0 |
| **11** | Vlad Nikolayevskyy, Katharina Kranzer, Stefan Niemann, Francis Drobniewski, Whole genome sequencing of Mycobacterium tuberculosis for detection of recent transmission and tracing outbreaks: A systematic review, Tuberculosis, Volume 98, May 2016, Pages 77-85, ISSN 1472-9792, http://dx.doi.org/10.1016/j.tube.2016.02.009. (http://www.sciencedirect.com/science/article/pii/S1472979215302705) Keywords: Tuberculosis; Epidemiology; Transmission; Next generation sequencing | no | yes | yes | yes | 3 | yes | yes | yes | yes | 4 | no | yes | yes | no | 2 | yes | yes | yes | yes | 4 |
| **12** | Mary C. Sheehan, Juleen Lam, Ana Navas-Acien, Howard H. Chang, Ambient air pollution epidemiology systematic review and meta-analysis: A review of reporting and methods practice, Environment International, Volumes 92–93, July–August 2016, Pages 647-656, ISSN 0160-4120, http://dx.doi.org/10.1016/j.envint.2016.02.016. (http://www.sciencedirect.com/science/article/pii/S0160412016300526) Keywords: Systematic review; Guidelines; Meta analysis; Environmental health; Air pollution; Climate change | no | yes | no | no | 1 | no | no | no | no | 0 | no | no | no | no | 0 | no | no | no | no | 0 |
| **13** | Francila Weidt Neiva, José Maria N. David, Regina Braga, Fernanda Campos, Towards pragmatic interoperability to support collaboration: A systematic review and mapping of the literature, Information and Software Technology, Volume 72, April 2016, Pages 137-150, ISSN 0950-5849, http://dx.doi.org/10.1016/j.infsof.2015.12.013. (http://www.sciencedirect.com/science/article/pii/S0950584916000021) Keywords: Pragmatic interoperability; Collaboration; Collaborative systems; Groupware; Interoperability | no | no | no | no | 0 | no | no | no | no | 0 | no | no | no | no | 0 | no | no | no | no | 0 |
| **14** | Colin B. Josephson, Sherry Sandy, Nathalie Jette, Tolulope T. Sajobi, Deborah Marshall, Samuel Wiebe, A systematic review of clinical decision rules for | no | no | no | no | 0 | no | no | no | no | 0 | no | no | no | no | 0 | no | no | no | no | 0 |
| | epilepsy, Epilepsy & Behavior, Volume 57, Part A, April 2016, Pages 69-76, ISSN 1525-5050, http://dx.doi.org/10.1016/j.yebeh.2016.01.019. (http://www.sciencedirect.com/science/article/pii/S1525505016000330) Keywords: Epilepsy; Diagnosis; Therapeutics; Rules; Clinical prediction | | | | | | | | | | | | | | | | | | | | |
| **15** | Domenico Ciliberto, Nicoletta Staropoli, Silvia Chiellino, Cirino Botta, Pierfrancesco Tassone, Pierosandro Tagliaferri, Systematic review and meta-analysis on targeted therapy in advanced pancreatic cancer, Pancreatology, Volume 16, Issue 2, March–April 2016, Pages 249-258, ISSN 1424-3903, http://dx.doi.org/10.1016/j.pan.2016.01.003. (http://www.sciencedirect.com/science/article/pii/S1424390316000089) Keywords: Advanced pancreatic cancer; Targeted therapy; Randomized clinical trials; Meta-analysis; Pathway; Pancreatic cancer | no | yes | no | no | 1 | no | no | no | no | 0 | no | no | no | no | 0 | no | no | no | no | 0 |
| **16** | Newton Spolaôr, Maria Carolina Monard, Grigorios Tsoumakas, Huei Diana Lee, A systematic review of multi-label feature selection and a new method based on label construction, Neurocomputing, Volume 180, 5 March 2016, Pages 3-15, ISSN 0925-2312, http://dx.doi.org/10.1016/j.neucom.2015.07.118. (http://www.sciencedirect.com/science/article/pii/S0925231215016197) Keywords: Feature ranking; Filter feature selection; Binary relevance; Information gain; Systematic review | yes | yes | no | no | 2 | no | no | no | no | 0 | no | no | no | no | 0 | no | no | no | no | 0 |
| **17** | Mohamed N. Saad, Mai S. Mabrouk, Ayman M. Eldeib, Olfat G. Shaker, Identification of rheumatoid arthritis biomarkers based on single nucleotide polymorphisms and haplotype blocks: A systematic review and meta-analysis, Journal of Advanced Research, Volume 7, Issue 1, January 2016, Pages 1-16, ISSN 2090-1232, http://dx.doi.org/10.1016/j.jare.2015.01.008. (http://www.sciencedirect.com/science/article/pii/S209012321500020X) Keywords: Haplotype block; Linkage disequilibrium; Major histocompatibility complex; Rheumatoid arthritis; Single nucleotide polymorphism | no | yes | yes | no | 2 | yes | yes | no | yes | 3 | no | no | no | no | 0 | no | no | no | no | 0 |
| **18** | Harapan Harapan, Cut Meurah Yeni, The role of microRNAs on angiogenesis and vascular pressure in preeclampsia: The evidence from systematic review, Egyptian Journal of Medical Human Genetics, Volume 16, Issue 4, October 2015, Pages 313-325, ISSN 1110-8630, http://dx.doi.org/10.1016/j.ejmhg.2015.03.006. (http://www.sciencedirect.com/science/article/pii/S1110863015000415) Keywords: | yes | yes | yes | yes | 4 | no | no | no | yes | 1 | no | no | no | no | 0 | no | no | no | no | 0 |
| | Preeclampsia pathogenesis; microRNA; miRNA; Angiogenesis; Systematic review | | | | | | | | | | | | | | | | | | | | |
| **19** | Taymour Mostafa, Mai Taymour, TNF-α −308 polymorphisms and male infertility risk: A meta-analysis and systematic review, Journal of Advanced Research, Volume 7, Issue 2, March 2016, Pages 185-192, ISSN 2090-1232, http://dx.doi.org/10.1016/j.jare.2015.07.001. (http://www.sciencedirect.com/science/article/pii/S2090123215000740) Keywords: Male infertility; TNF-α; Polymorphism; Meta-analysis; Semen; Sperms | yes | yes | yes | no | 3 | no | no | no | yes | 1 | no | no | no | no | 0 | no | no | no | no | 0 |
| **20** | Steven Habbous, Vincent Pang, Wei Xu, Eitan Amir, Geoffrey Liu, Human papillomavirus and host genetic polymorphisms in carcinogenesis: A systematic review and meta-analysis, Journal of Clinical Virology, Volume 61, Issue 2, October 2014, Pages 220-229, ISSN 1386-6532, http://dx.doi.org/10.1016/j.jcv.2014.07.019. (http://www.sciencedirect.com/science/article/pii/S1386653214002996) Keywords: Polymorphism; Human papillomavirus; Cancer; Interaction; Cervix; Head and neck | yes | yes | yes | yes | 4 | yes | yes | yes | yes | 4 | no | no | no | no | 0 | no | no | no | no | 0 |
| **21** | Juliessa M. Pavon, Soheir S. Adam, Zayd A. Razouki, Jennifer R. McDuffie, Paul F. Lachiewicz, Andrzej S. Kosinski, Christopher A. Beadles, Thomas L. Ortel, Avishek Nagi, John W. Williams Jr., Effectiveness of Intermittent Pneumatic Compression Devices for Venous Thromboembolism Prophylaxis in High-Risk Surgical Patients: A Systematic Review, The Journal of Arthroplasty, Volume 31, Issue 2, February 2016, Pages 524-532, ISSN 0883-5403, http://dx.doi.org/10.1016/j.arth.2015.09.043. (http://www.sciencedirect.com/science/article/pii/S0883540315008815) Keywords: intermittent pneumatic compression devices; venous thromboembolism prophylaxis; anticoagulation; joint arthroplasty; surgery | no | no | no | no | 0 | no | no | no | no | 0 | no | no | no | no | 0 | no | no | no | no | 0 |
| **22** | Freija Verdoodt, Søren Friis, Christian Dehlendorff, Vanna Albieri, Susanne K. Kjaer, Non-steroidal anti-inflammatory drug use and risk of endometrial cancer: A systematic review and meta-analysis of observational studies, Gynecologic Oncology, Volume 140, Issue 2, February 2016, Pages 352-358, ISSN 0090-8258, http://dx.doi.org/10.1016/j.ygyno.2015.12.009. (http://www.sciencedirect.com/science/article/pii/S0090825815302134) Keywords: Pharmacoepidemiology; Endometrial cancer; Non-steroidal anti-inflammatory drugs; Aspirin; Cancer prevention | no | no | no | no | 0 | no | no | no | no | 0 | no | no | no | no | 0 | no | no | no | no | 0 |
| **23** | Amirhossein Sahebkar, Maria-Corina Serban, Dimitri P. Mikhailidis, Peter P. Toth, Paul Muntner, Sorin Ursoniu, Svetlana Mosterou, Stephen Glasser, Seth S. Martin, Steven R. Jones, Manfredi Rizzo, Jacek Rysz, Allan D. Sniderman, Michael J. Pencina, Maciej Banach, Head-to-head comparison of statins versus fibrates in reducing plasma fibrinogen concentrations: A systematic review and meta-analysis, Pharmacological Research, Volume 103, January 2016, Pages 236-252, ISSN 1043-6618, http://dx.doi.org/10.1016/j.phrs.2015.12.001. (http://www.sciencedirect.com/science/article/pii/S1043661815302085) Keywords: Fibrates; Statins; Fibrinogen; Meta-analysis; Inflammation | no | no | no | no | 0 | no | no | no | no | 0 | no | no | no | no | 0 | no | no | no | no | 0 |
| **24** | Xuexian Fang, Jiayu Wei, Xuyan He, Peng An, Hao Wang, Li Jiang, Dandan Shao, Han Liang, Yi Li, Fudi Wang, Junxia Min, Landscape of dietary factors associated with risk of gastric cancer: A systematic review and dose-response meta-analysis of prospective cohort studies, European Journal of Cancer, Volume 51, Issue 18, December 2015, Pages 2820-2832, ISSN 0959-8049, http://dx.doi.org/10.1016/j.ejca.2015.09.010. (http://www.sciencedirect.com/science/article/pii/S0959804915008564) Keywords: Gastric cancer; Nutrition; Salt; Alcohol; Prospective; Dose-response; Meta-analysis | no | no | no | no | 0 | no | no | no | no | 0 | no | no | no | no | 0 | no | no | no | no | 0 |
| **25** | M.J. Bown, R.D. Sayers, Invited Commentary on “Potential Circulating Biomarkers for Abdominal Aortic Aneurysm Expansion and Rupture - a Systematic Review” by Urbonavicius et al., European Journal of Vascular and Endovascular Surgery, Volume 36, Issue 3, September 2008, Pages 281-282, ISSN 1078-5884, http://dx.doi.org/10.1016/j.ejvs.2008.06.007. (http://www.sciencedirect.com/science/article/pii/S1078588408003213) | no | no | no | no | 0 | no | no | no | no | 0 | no | no | no | no | 0 | no | no | no | no | 0 |
| **26** | Zohreh Dehghani Champiri, Seyed Reza Shahamiri, Siti Salwah Binti Salim, A systematic review of scholar context-aware recommender systems, Expert Systems with Applications, Volume 42, Issue 3, 15 February 2015, Pages 1743-1758, ISSN 0957-4174, http://dx.doi.org/10.1016/j.eswa.2014.09.017. (http://www.sciencedirect.com/science/article/pii/S0957417414005569) Keywords: Context-aware recommender system; Context-awareness; Contextual information; Academic digital library | no | no | no | no | 0 | no | no | no | no | 0 | no | no | no | no | 0 | no | no | no | no | 0 |
| **27** | Susan Kühnast, Marta Fiocco, José W.A. van der Hoorn, Hans M.G. Princen, J. Wouter Jukema, Innovative pharmaceutical interventions in cardiovascular disease: | no | no | no | no | 0 | no | no | no | no | 0 | no | no | no | no | 0 | no | no | no | no | 0 |
| | Focusing on the contribution of non-HDL-C/LDL-C-lowering versus HDL-C-raisingA systematic review and meta-analysis of relevant preclinical studies and clinical trials, European Journal of Pharmacology, Volume 763, Part A, 15 September 2015, Pages 48-63, ISSN 0014-2999, http://dx.doi.org/10.1016/j.ejphar.2015.03.089. (http://www.sciencedirect.com/science/article/pii/S0014299915004501) Index terms: Cardiovascular disease; Clinical outcome; Myocardial infarction; LDL-cholesterol; Non-HDL-cholesterol; HDL-cholesterol; Atherosclerosis; Clinical trials; APOE*3Leiden.CETP mice; Mouse; Rabbit; Hamster; Niacin; Fibrates; Glitazones; PPAR agonists; CETP inhibition; SR-BI inhibitor; ABCA1 degradation inhibitors; Reconstituted HDL; Delipidated HDL; Apolipoprotein A-I Milano; Apolipoprotein A-I mimetic; Apolipoprotein A-I inducer; LCAT; HDL-C-raising pharmaceutical interventions; Preclinical studies; Systematic review; Randomised controlled trials; Meta-analysis | | | | | | | | | | | | | | | | | | | | |
| **28** | Harsha Moole, Anushi Shah, Raghuveer R. Boddireddy, Vishnu Moole, Achuta Uppu, Sa1589 Comparing the Efficacy of Norepinephrine Plus Albumin Versus Terlipressin Plus Albumin in Treating Hepatorenal Syndrome: A Systematic Review and Meta-Analysis, Gastroenterology, Volume 150, Issue 4, Supplement 1, April 2016, Page S335, ISSN 0016-5085, http://dx.doi.org/10.1016/S0016-5085(16)31179-9. (http://www.sciencedirect.com/science/article/pii/S0016508516311799) | no | no | no | no | 0 | no | no | no | no | 0 | no | no | no | no | 0 | no | no | no | no | 0 |
| **29** | Madhusmita Behera, Rathi N. Pillai, Taofeek K. Owonikoko, Sungjin Kim, Conor Steuer, Zhengjia Chen, Nabil F. Saba, Chandra P. Belani, Fadlo R. Khuri, Suresh S. Ramalingam, Bevacizumab in Combination with Taxane versus Non-Taxane Containing Regimens for Advanced/Metastatic Nonsquamous Non–Small-Cell Lung Cancer: A Systematic Review, Journal of Thoracic Oncology, Volume 10, Issue 8, August 2015, Pages 1142-1147, ISSN 1556-0864, http://dx.doi.org/10.1097/JTO.0000000000000572. (http://www.sciencedirect.com/science/article/pii/S1556086415335401) Key Words: bevacizumab; taxanes; systematic review | no | no | no | no | 0 | no | no | no | no | 0 | no | no | no | no | 0 | no | no | no | no | 0 |
| **30** | Amir Khoshbin, Timothy Leroux, David Wasserstein, Paul Marks, John Theodoropoulos, Darrell Ogilvie-Harris, Rajiv Gandhi, Kirat Takhar, Grant Lum, Jaskarndip Chahal, The Efficacy of Platelet-Rich Plasma in the Treatment of | no | no | no | no | 0 | no | no | no | no | 0 | no | no | no | no | 0 | no | no | no | no | 0 |
| | Symptomatic Knee Osteoarthritis: A Systematic Review With Quantitative Synthesis, Arthroscopy: The Journal of Arthroscopic & Related Surgery, Volume 29, Issue 12, December 2013, Pages 2037-2048, ISSN 0749-8063, http://dx.doi.org/10.1016/j.arthro.2013.09.006. (http://www.sciencedirect.com/science/article/pii/S0749806313010050) | | | | | | | | | | | | | | | | | | | | |
| **31** | Sreekanth Vemulapalli, Rowena J. Dolor, Vic Hasselblad, Kristine Schmit, Adam Banks, Brooke Heidenfelder, Manesh R. Patel, W. Schuyler Jones, Supervised vs unsupervised exercise for intermittent claudication: A systematic review and meta-analysis, American Heart Journal, Volume 169, Issue 6, June 2015, Pages 924-937.e3, ISSN 0002-8703, http://dx.doi.org/10.1016/j.ahj.2015.03.009. (http://www.sciencedirect.com/science/article/pii/S0002870315001799) | no | no | no | no | 0 | no | no | no | no | 0 | no | no | no | no | 0 | no | no | no | no | 0 |
| **32** | Daniël A Korevaar, Guus A Westerhof, Junfeng Wang, Jérémie F Cohen, René Spijker, Peter J Sterk, Elisabeth H Bel, Patrick M M Bossuyt, Diagnostic accuracy of minimally invasive markers for detection of airway eosinophilia in asthma: a systematic review and meta-analysis, The Lancet Respiratory Medicine, Volume 3, Issue 4, April 2015, Pages 290-300, ISSN 2213-2600, http://dx.doi.org/10.1016/S2213-2600(15)00050-8. (http://www.sciencedirect.com/science/article/pii/S2213260015000508) | no | no | no | no | 0 | no | no | no | no | 0 | no | no | no | no | 0 | no | no | no | no | 0 |
| **33** | J. Zhang, Y. Deng, Y. Cai, H. Hu, J. Xiao, J. Ling, 2101 FOLFOXIRI versus FOLFOX/FOLFIRI plus Bevacizumab as first-line treatment in metastatic colorectal cancer: A systematic review and network Meta-analysis of phase III study, European Journal of Cancer, Volume 51, Supplement 3, September 2015, Pages S362-S363, ISSN 0959-8049, http://dx.doi.org/10.1016/S0959-8049(16)31023-1. (http://www.sciencedirect.com/science/article/pii/S0959804916310231) | no | no | no | no | 0 | no | no | no | no | 0 | no | no | no | no | 0 | no | no | no | no | 0 |
| **34** | D. Cho, F. Roncolato, J.J. Man, S.J. Lord, M. Links, J. Yeses, C.K. Lee, 350 How effective are novel biological therapies? A systematic review of 121 randomized controlled trials (RCTs), European Journal of Cancer, Volume 51, Supplement 3, September 2015, Pages S70-S71, ISSN 0959-8049, http://dx.doi.org/10.1016/S0959-8049(16)30213-1. (http://www.sciencedirect.com/science/article/pii/S0959804916302131) | no | no | no | no | 0 | no | no | no | no | 0 | no | no | no | no | 0 | no | no | no | no | 0 |
| **35** | Anne J Breugom, Marloes Swets, Jean-François Bosset, Laurence Collette, Aldo | no | no | no | no | 0 | no | no | no | no | 0 | no | no | no | no | 0 | no | no | no | no | 0 |
| | Sainato, Luca Cionini, Rob Glynne-Jones, Nicholas Counsell, Esther Bastiaannet, Colette B M van den Broek, Gerrit-Jan Liefers, Hein Putter, Cornelis J H van de Velde, Adjuvant chemotherapy after preoperative (chemo)radiotherapy and surgery for patients with rectal cancer: a systematic review and meta-analysis of individual patient data, The Lancet Oncology, Volume 16, Issue 2, February 2015, Pages 200-207, ISSN 1470-2045, http://dx.doi.org/10.1016/S1470-2045(14)71199-4. (http://www.sciencedirect.com/science/article/pii/S1470204514711994) | | | | | | | | | | | | | | | | | | | | |
| **36** | Kelly S. Acharya, Meredith P. Provost, Jason S. Yeh, Chaitanya R Acharya, Suheil J. Muasher, Ectopic pregnancy rates in frozen versus fresh embryo transfer in in vitro fertilization: A systematic review and meta-analysis, Middle East Fertility Society Journal, Volume 19, Issue 4, December 2014, Pages 233-238, ISSN 1110-5690, http://dx.doi.org/10.1016/j.mefs.2014.09.001. (http://www.sciencedirect.com/science/article/pii/S1110569014200173) KEYWORDS: Ectopic; Fresh; Frozen; Embryo; Meta-analysis | no | no | no | no | 0 | no | no | no | no | 0 | no | no | no | no | 0 | no | no | no | no | 0 |
| **37** | Patrick P.M. Schrama, Martijn S. Stenneberg, Cees Lucas, Emiel van Trijffel, Intraexaminer Reliability of Hand-Held Dynamometry in the Upper Extremity: A Systematic Review, Archives of Physical Medicine and Rehabilitation, Volume 95, Issue 12, December 2014, Pages 2444-2469, ISSN 0003-9993, http://dx.doi.org/10.1016/j.apmr.2014.05.019. (http://www.sciencedirect.com/science/article/pii/S0003999314004146) Keywords: Extremities; Isometric contraction; Muscle strength dynamometer; Observer variation; Rehabilitation; Reproducibility of results | no | no | no | no | 0 | no | no | no | no | 0 | no | no | no | no | 0 | no | no | no | no | 0 |
| **38** | Álvaro Machado Dias, Carlos Gustavo Mansur, Martin Myczkowski, Marco Marcolin, Whole field tendencies in transcranial magnetic stimulation: A systematic review with data and text mining, Asian Journal of Psychiatry, Volume 4, Issue 2, June 2011, Pages 107-112, ISSN 1876-2018, http://dx.doi.org/10.1016/j.ajp.2011.03.003. (http://www.sciencedirect.com/science/article/pii/S1876201811000372) Keywords: TMS; Neuropsychiatry; Text mining; Systematic review; Depression | yes | no | no | no | 1 | no | no | no | no | 0 | no | no | no | no | 0 | no | no | no | no | 0 |
| **39** | Charline Maertens de Noordhout, Brecht Devleesschauwer, Frederick J Angulo, Geert Verbeke, Juanita Haagsma, Martyn Kirk, Arie Havelaar, Niko Speybroeck, The global burden of listeriosis: a systematic review and meta-analysis, The Lancet | no | no | no | no | 0 | no | no | no | no | 0 | no | no | no | no | 0 | no | no | no | no | 0 |
| | Infectious Diseases, Volume 14, Issue 11, November 2014, Pages 1073-1082, ISSN 1473-3099, http://dx.doi.org/10.1016/S1473-3099(14)70870-9. (http://www.sciencedirect.com/science/article/pii/S1473309914708709) | | | | | | | | | | | | | | | | | | | | |
| **40** | Wynanda A. van Enst, Rob J.P.M. Scholten, Penny Whiting, Aeilko H. Zwinderman, Lotty Hooft, Meta-epidemiologic analysis indicates that MEDLINE searches are sufficient for diagnostic test accuracy systematic reviews, Journal of Clinical Epidemiology, Volume 67, Issue 11, November 2014, Pages 1192-1199, ISSN 0895-4356, http://dx.doi.org/10.1016/j.jclinepi.2014.05.008. (http://www.sciencedirect.com/science/article/pii/S0895435614002005) Keywords: Diagnostic test accuracy; Meta-analyses; Dissemination bias; Comprehensive searches; MEDLINE; Systematic review | no | no | no | no | 0 | no | no | no | no | 0 | no | no | no | no | 0 | no | no | no | no | 0 |
| **41** | Christiana A. Naaktgeboren, Wynanda A. van Enst, Eleanor A. Ochodo, Joris A.H. de Groot, Lotty Hooft, Mariska M. Leeflang, Patrick M. Bossuyt, Karel G.M. Moons, Johannes B. Reitsma, Systematic overview finds variation in approaches to investigating and reporting on sources of heterogeneity in systematic reviews of diagnostic studies, Journal of Clinical Epidemiology, Volume 67, Issue 11, November 2014, Pages 1200-1209, ISSN 0895-4356, http://dx.doi.org/10.1016/j.jclinepi.2014.05.018. (http://www.sciencedirect.com/science/article/pii/S0895435614002297) Keywords: Meta-analysis; Diagnostic techniques and procedures/standards; Sensitivity and specificity; Data interpretation; Statistical; Bias (epidemiology) | no | no | no | no | 0 | no | no | no | no | 0 | no | no | no | no | 0 | no | no | no | no | 0 |
| **42** | Julie B. Hebnes, Tina B. Olesen, Anne Katrine Duun-Henriksen, Christian Munk, Bodil Norrild, Susanne K. Kjaer, Prevalence of Genital Human Papillomavirus among Men in Europe: Systematic Review and Meta-Analysis, The Journal of Sexual Medicine, Volume 11, Issue 11, November 2014, Pages 2630-2644, ISSN 1743-6095, http://dx.doi.org/10.1111/jsm.12652. (http://www.sciencedirect.com/science/article/pii/S174360951530607X) Keywords: Human Papillomavirus; Genital; Men; Europe; Systematic Review; Meta-Analysis | no | no | no | no | 0 | no | no | no | no | 0 | no | no | no | no | 0 | no | no | no | no | 0 |
| **43** | Madhusmita Behera, Taofeek K. Owonikoko, Sungjin Kim, Zhengjia Chen, Kristin Higgins, Suresh S. Ramalingam, Dong M. Shin, Fadlo R. Khuri, Jonathan J. Beitler, Nabil F. Saba, Concurrent therapy with taxane versus non-taxane containing | no | no | no | no | 0 | no | no | no | no | 0 | no | no | no | no | 0 | no | no | no | no | 0 |
| | regimens in locally advanced squamous cell carcinomas of the head and neck (SCCHN): A systematic review, Oral Oncology, Volume 50, Issue 9, September 2014, Pages 888-894, ISSN 1368-8375, http://dx.doi.org/10.1016/j.oraloncology.2014.06.014. (http://www.sciencedirect.com/science/article/pii/S1368837514001730) Keywords: Concurrent therapy in head and neck cancer; Taxol in head and neck cancer; Taxanes in concurrent regimens for head and neck cancer | | | | | | | | | | | | | | | | | | | | |
| **44** | Julia Ratter, Lorenz Radlinger, Cees Lucas, Several submaximal exercise tests are reliable, valid and acceptable in people with chronic pain, fibromyalgia or chronic fatigue: a systematic review, Journal of Physiotherapy, Volume 60, Issue 3, September 2014, Pages 144-150, ISSN 1836-9553, http://dx.doi.org/10.1016/j.jphys.2014.06.011. (http://www.sciencedirect.com/science/article/pii/S1836955314000794) Keywords: Chronic pain; Fatigue syndrome (chronic); Fibromyalgia; Exercise test; Psychometrics; Review (publication type) | no | no | no | no | 0 | no | no | no | no | 0 | no | no | no | no | 0 | no | no | no | no | 0 |
| **45** | M.O. Needham, D. Paech, S. Norris, S. Campbell, PIN2 Assessing the Diagnostic Performance of Human Immunodeficiency Virus Type 1 (HIV-1) Tropism Testing Methods: A Systematic Review of Genotypic Sequencing of the Third Hypervariable (V3) Loop and Enhanced-Sensitivity Phenotypic Assay Technology, Value in Health, Volume 15, Issue 7, November 2012, Pages A665-A666, ISSN 1098-3015, http://dx.doi.org/10.1016/j.jval.2012.08.371. (http://www.sciencedirect.com/science/article/pii/S1098301512020840) | no | no | no | yes | 1 | no | no | no | no | 0 | no | no | no | no | 0 | no | no | no | no | 0 |
| **46** | Maartje N. Niemeijer, Iris J. Grooten, Nikki Vos, Joke M.J. Bais, Joris A. van der Post, Ben W. Mol, Tessa J. Roseboom, Mariska M.G. Leeflang, Rebecca C. Painter, Diagnostic markers for hyperemesis gravidarum: a systematic review and metaanalysis, American Journal of Obstetrics and Gynecology, Volume 211, Issue 2, August 2014, Pages 150.e1-150.e15, ISSN 0002-9378, http://dx.doi.org/10.1016/j.ajog.2014.02.012. (http://www.sciencedirect.com/science/article/pii/S0002937814001392) Key words: biomarker; diagnosis; hyperemesis gravidarum; nausea and vomiting; pregnancy | no | no | no | no | 0 | no | no | no | no | 0 | no | no | no | no | 0 | no | no | no | no | 0 |
| **47** | Tina Bech Olesen, Malene Frøsig Svahn, Mette Tuxen Faber, Anne Katrine Duun-Henriksen, Jette Junge, Bodil Norrild, Susanne K. Kjaer, Prevalence of Human | no | no | no | no | 0 | no | no | no | no | 0 | no | no | no | no | 0 | no | no | no | no | 0 |
| | Papillomavirus in endometrial cancer: A systematic review and meta-analysis, Gynecologic Oncology, Volume 134, Issue 1, July 2014, Pages 206-215, ISSN 0090-8258, http://dx.doi.org/10.1016/j.ygyno.2014.02.040. (http://www.sciencedirect.com/science/article/pii/S0090825814002042) Keywords: Endometrial cancer; Human Papillomavirus; Prevalence; Systematic review; Meta-analysis | | | | | | | | | | | | | | | | | | | | |
| **48** | Louise Baandrup, Louise T. Thomsen, Tina Bech Olesen, Klaus Kaae Andersen, Bodil Norrild, Susanne K. Kjaer, The prevalence of human papillomavirus in colorectal adenomas and adenocarcinomas: A systematic review and meta-analysis, European Journal of Cancer, Volume 50, Issue 8, May 2014, Pages 1446-1461, ISSN 0959-8049, http://dx.doi.org/10.1016/j.ejca.2014.01.019. (http://www.sciencedirect.com/science/article/pii/S0959804914000926) Keywords: Colorectal cancer; Human papillomavirus; Prevalence; Metaanalysis | no | no | no | no | 0 | no | no | no | no | 0 | no | no | no | no | 0 | no | no | no | no | 0 |
| **49** | Remy R. Coeytaux, Kristine M. Schmit, Bryan D. Kraft, Andrzej S. Kosinski, Alicea M. Mingo, Lisa M. Vann, Daniel L. Gilstrap, C. William Hargett, Brooke Heidenfelder, Rowena J. Dolor, Douglas C. McCrory, Comparative Effectiveness and Safety of Drug Therapy for Pulmonary Arterial Hypertension: A Systematic Review and Meta-analysis, Chest, Volume 145, Issue 5, May 2014, Pages 1055-1063, ISSN 0012-3692, http://dx.doi.org/10.1378/chest.13-1864. (http://www.sciencedirect.com/science/article/pii/S0012369215345980) | no | no | no | no | 0 | no | no | no | no | 0 | no | no | no | no | 0 | no | no | no | no | 0 |
| **50** | Annemarie Coolbrandt, Hans Wildiers, Bert Aertgeerts, Elisa Van der Elst, Annouschka Laenen, Bernadette Dierckx de Casterlé, Theo van Achterberg, Koen Milisen, Characteristics and effectiveness of complex nursing interventions aimed at reducing symptom burden in adult patients treated with chemotherapy: A systematic review of randomized controlled trials, International Journal of Nursing Studies, Volume 51, Issue 3, March 2014, Pages 495-510, ISSN 0020-7489, http://dx.doi.org/10.1016/j.ijnurstu.2013.08.008. (http://www.sciencedirect.com/science/article/pii/S0020748913002484) Keywords: Cancer; Chemotherapy; Intervention; Complex interventions; Nursing; Symptom burden; Side effects; Systematic review | no | no | no | no | 0 | no | no | no | no | 0 | no | no | no | no | 0 | no | no | no | no | 0 |
| **51** | M.M.G. Leeflang, Systematic reviews and meta-analyses of diagnostic test accuracy, Clinical Microbiology and Infection, Volume 20, Issue 2, February 2014, | no | no | no | no | 0 | no | no | no | no | 0 | no | no | no | no | 0 | no | no | no | no | 0 |
| | Pages 105-113, ISSN 1198-743X, http://dx.doi.org/10.1111/1469-0691.12474. (http://www.sciencedirect.com/science/article/pii/S1198743X14600434) Keywords: diagnosis; diagnostic test accuracy; evidence-based medicine; meta-analyses; sensitivity and specificity; systematic reviews | | | | | | | | | | | | | | | | | | | | |
| **52** | Kristine M. Schmit, Remy R. Coeytaux, Adam P. Goode, Douglas C. McCrory, William S. Yancy Jr, Alex R. Kemper, Vic Hasselblad, Brooke L. Heidenfelder, Gillian D. Sanders, Evaluating Cough Assessment Tools: A Systematic Review, Chest, Volume 144, Issue 6, December 2013, Pages 1819-1826, ISSN 0012-3692, http://dx.doi.org/10.1378/chest.13-0310. (http://www.sciencedirect.com/science/article/pii/S0012369215486921) | no | no | no | no | 0 | no | no | no | no | 0 | no | no | no | no | 0 | no | no | no | no | 0 |
| **53** | William S. Yancy Jr, Douglas C. McCrory, Remy R. Coeytaux, Kristine M. Schmit, Alex R. Kemper, Adam Goode, Victor Hasselblad, Brooke L. Heidenfelder, Gillian D. Sanders, Efficacy and Tolerability of Treatments for Chronic Cough: A Systematic Review and Meta-analysis, Chest, Volume 144, Issue 6, December 2013, Pages 1827-1838, ISSN 0012-3692, http://dx.doi.org/10.1378/chest.13-0490. (http://www.sciencedirect.com/science/article/pii/S0012369215486933) | no | no | no | no | 0 | no | no | no | no | 0 | no | no | no | no | 0 | no | no | no | no | 0 |
| **54** | Zheng Li, He Zhang, Liam O'Brien, Rainbow Cai, Shayne Flint, On evaluating commercial Cloud services: A systematic review, Journal of Systems and Software, Volume 86, Issue 9, September 2013, Pages 2371-2393, ISSN 0164-1212, http://dx.doi.org/10.1016/j.jss.2013.04.021. (http://www.sciencedirect.com/science/article/pii/S0164121213000915) Keywords: Cloud Computing; Cloud service evaluation; Systematic literature review | no | no | no | no | 0 | no | no | no | no | 0 | no | no | no | no | 0 | no | no | no | no | 0 |
| **55** | Sébastien P.J. Krul, Antoine H.G. Driessen, Aeilko H. Zwinderman, Wim J. van Boven, Arthur A.M. Wilde, Jacques M.T. de Bakker, Joris R. de Groot, Navigating the mini-maze: Systematic review of the first results and progress of minimally-invasive surgery in the treatment of atrial fibrillation, International Journal of Cardiology, Volume 166, Issue 1, 5 June 2013, Pages 132-140, ISSN 0167-5273, http://dx.doi.org/10.1016/j.ijcard.2011.10.011. (http://www.sciencedirect.com/science/article/pii/S0167527311018699) Keywords: Ablation; Atrial fibrillation; Minimally-invasive surgery | no | no | no | no | 0 | no | no | no | no | 0 | no | no | no | no | 0 | no | no | no | no | 0 |
| **56** | Philip J. Leisy, Remy R. Coeytaux, Galen S. Wagner, Eugene H. Chung, Amanda J. McBroom, Cynthia L. Green, John W. Williams Jr., Gillian D. Sanders, ECG-based | no | no | no | no | 0 | no | no | no | no | 0 | no | no | no | no | 0 | no | no | no | no | 0 |
| | signal analysis technologies for evaluating patients with acute coronary syndrome: A systematic review, Journal of Electrocardiology, Volume 46, Issue 2, March–April 2013, Pages 92-97, ISSN 0022-0736, http://dx.doi.org/10.1016/j.jelectrocard.2012.11.010. (http://www.sciencedirect.com/science/article/pii/S0022073612004815) Keywords: Electrocardiography; Acute coronary syndrome; ECG-based signal analysis | | | | | | | | | | | | | | | | | | | | |
| **57** | W. Peter Bekkering, Theodora P.M. Vliet Vlieland, Marta Fiocco, Hendrik M. Koopman, Jan W. Schoones, Rob G.H.H. Nelissen, Antonie H.M. Taminiau, Quality of life, functional ability and physical activity after different surgical interventions for bone cancer of the leg: A systematic review, Surgical Oncology, Volume 21, Issue 2, June 2012, Pages e39-e47, ISSN 0960-7404, http://dx.doi.org/10.1016/j.suronc.2011.09.002. (http://www.sciencedirect.com/science/article/pii/S0960740411000739) Keywords: Systematic review; Bone tumours; Lower extremity; Quality of life; Functional ability; Physical activity | no | no | no | no | 0 | no | no | no | no | 0 | no | no | no | no | 0 | no | no | no | no | 0 |
| **58** | An-Ting T. Lu, Shelley R. Salpeter, Anthony E. Reeve, Steven Eschrich, Patrick G. Johnston, Alain J. Barrier, Francois Bertucci, Nicholas S. Buckley, Edwin E. Salpeter, Albert Y. Lin, Gene Expression Profiles as Predictors of Poor Outcomes in Stage II Colorectal Cancer: A Systematic Review and Meta-analysis, Clinical Colorectal Cancer, Volume 8, Issue 4, October 2009, Pages 207-214, ISSN 1533-0028, http://dx.doi.org/10.3816/CCC.2009.n.035. (http://www.sciencedirect.com/science/article/pii/S1533002811703747) Keywords: DNA microarray; Gene signature; Likelihood ratio; Prognosis prediction | no | no | yes | yes | 2 | yes | no | no | yes | 2 | no | no | no | no | 0 | no | no | no | no | 0 |
| **59** | C. Jimenez-Rivera, K.S. Jolin-Dahel, K.J. Fortinsky, P. Gozdyra, E.I. Benchimol, 1387 INTERNATIONAL INCIDENCE AND OUTCOMES OF BILIARY ATRESIA: A SYSTEMATIC REVIEW OF POPULATION-BASED STUDIES, Journal of Hepatology, Volume 56, Supplement 2, April 2012, Pages S544-S545, ISSN 0168-8278, http://dx.doi.org/10.1016/S0168-8278(12)61398-1. (http://www.sciencedirect.com/science/article/pii/S0168827812613981) | no | no | no | no | 0 | no | no | no | no | 0 | no | no | no | no | 0 | no | no | no | no | 0 |
| **60** | Aaron M. Cohen, Kyle Ambert, Marian McDonagh, Cross-Topic Learning for Work Prioritization in Systematic Review Creation and Update, Journal of the American Medical Informatics Association, Volume 16, Issue 5, September– | no | no | no | yes | 1 | no | no | no | no | 0 | no | no | no | no | 0 | no | no | no | no | 0 |
| | October 2009, Pages 690-704, ISSN 1067-5027, http://dx.doi.org/10.1197/jamia.M3162. (http://www.sciencedirect.com/science/article/pii/S1067502709001224) | | | | | | | | | | | | | | | | | | | | |
| **61** | J. Rauw, M. Ennis, M. Krzyzanowska, S. Sridhar, 269 Does the Addition of Molecular Targeted Therapy to Standard Treatments Lead to Better or Worse Outcomes Overall? A Systematic Review and Meta-analysis of EGFR-targeted Therapies Used in Combination with Standard Treatments, European Journal of Cancer, Volume 48, Supplement 6, November 2012, Pages 82-83, ISSN 0959-8049, http://dx.doi.org/10.1016/S0959-8049(12)72067-1. (http://www.sciencedirect.com/science/article/pii/S0959804912720671) | yes | no | no | no | 1 | no | no | no | no | 0 | no | no | no | no | 0 | no | no | no | no | 0 |
| **62** | Bianca M. Buurman, Barbara C. van Munster, Johanna C. Korevaar, Rob J. de Haan, Sophia E. de Rooij, Variability in measuring (instrumental) activities of daily living functioning and functional decline in hospitalized older medical patients: a systematic review, Journal of Clinical Epidemiology, Volume 64, Issue 6, June 2011, Pages 619-627, ISSN 0895-4356, http://dx.doi.org/10.1016/j.jclinepi.2010.07.005. (http://www.sciencedirect.com/science/article/pii/S0895435610002611) Keywords: Functional decline; Systematic review; Measurement; Older patient; Hospitalized; Activities of daily living; Instrumental activities of daily living | no | no | no | no | 0 | no | no | no | no | 0 | no | no | no | no | 0 | no | no | no | no | 0 |
| **63** | Anna M. Musters, Sjoerd Repping, Johanna C. Korevaar, Sebastiaan Mastenbroek, Jacqueline Limpens, Fulco van der Veen, Mariëtte Goddijn, Pregnancy outcome after preimplantation genetic screening or natural conception in couples with unexplained recurrent miscarriage: a systematic review of the best available evidence, Fertility and Sterility, Volume 95, Issue 6, May 2011, Pages 2153-2157.e3, ISSN 0015-0282, http://dx.doi.org/10.1016/j.fertnstert.2010.12.022. (http://www.sciencedirect.com/science/article/pii/S001502821002964X) Key Words: Habitual abortion; preimplantation genetic screening; live birth rates; miscarriage rates; natural conception | no | yes | no | yes | 2 | no | no | no | no | 0 | no | no | no | no | 0 | no | no | no | no | 0 |
| **64** | M. Glogger, P. Schnell-Inderst, A. Luzak, A.. Krueger, U. Siebert, PCN277 - New Reimbursement Schemes for Stratified Medicine in Oncology – A Systematic Review, Value in Health, Volume 17, Issue 7, November 2014, Page A663, ISSN 1098-3015, http://dx.doi.org/10.1016/j.jval.2014.08.2436. | no | no | no | no | 0 | no | no | no | no | 0 | no | no | no | no | 0 | no | no | no | no | 0 |
| | (http://www.sciencedirect.com/science/article/pii/S1098301514043666) | | | | | | | | | | | | | | | | | | | | |
| **65** | Marsha van Leeuwen, Brent C. Opmeer, Yildirim Yilmaz, Jacqueline Limpens, Mireille J. Serlie, Ben Willem J. Mol, Accuracy of the random glucose test as screening test for gestational diabetes mellitus: a systematic review, European Journal of Obstetrics & Gynecology and Reproductive Biology, Volume 154, Issue 2, February 2011, Pages 130-135, ISSN 0301-2115, http://dx.doi.org/10.1016/j.ejogrb.2010.11.002. (http://www.sciencedirect.com/science/article/pii/S0301211510005464) Keywords: Gestational diabetes mellitus; Random glucose test; Screening | no | no | no | no | 0 | no | no | no | no | 0 | no | no | no | no | 0 | no | no | no | no | 0 |
| **66** | Sally W. Aboelela, Lisa Saiman, Patricia Stone, Franklin D. Lowy, Dave Quiros, Elaine Larson, Effectiveness of barrier precautions and surveillance cultures to control transmission of multidrug-resistant organisms: A systematic review of the literature, American Journal of Infection Control, Volume 34, Issue 8, October 2006, Pages 484-494, ISSN 0196-6553, http://dx.doi.org/10.1016/j.ajic.2006.03.008. (http://www.sciencedirect.com/science/article/pii/S0196655306002380) | no | no | no | no | 0 | no | no | no | no | 0 | no | no | no | no | 0 | no | no | no | no | 0 |
| **67** | Lineke M. Tak, Harriëtte Riese, Geertruida H. de Bock, Andiappan Manoharan, Iris C. Kok, Judith G.M. Rosmalen, As good as it gets? A meta-analysis and systematic review of methodological quality of heart rate variability studies in functional somatic disorders, Biological Psychology, Volume 82, Issue 2, October 2009, Pages 101-110, ISSN 0301-0511, http://dx.doi.org/10.1016/j.biopsycho.2009.05.002. (http://www.sciencedirect.com/science/article/pii/S0301051109001008) Keywords: Autonomic nervous system; Heart rate variability; Parasympathetic nervous system; Stress; Functional somatic disorders; Somatoform | no | no | no | no | 0 | no | no | no | no | 0 | no | no | no | no | 0 | no | no | no | no | 0 |
| **68** | J.A.M. Bramer, J.H. van Linge, R.J. Grimer, R.J.P.M. Scholten, Prognostic factors in localized extremity osteosarcoma: A systematic review, European Journal of Surgical Oncology (EJSO), Volume 35, Issue 10, October 2009, Pages 1030-1036, ISSN 0748-7983, http://dx.doi.org/10.1016/j.ejso.2009.01.011. (http://www.sciencedirect.com/science/article/pii/S0748798309000286) Keywords: Humans; Osteosarcoma; Bone neoplasms; Prognosis; Survival analysis; Extremities | no | no | no | no | 0 | no | no | no | no | 0 | no | no | no | no | 0 | no | no | no | no | 0 |
| **69** | Wouter L. Curvers, Frank J.C. van den Broek, Johannes B. Reitsma, Evelien Dekker, Jacques J.G.H.M. Bergman, Systematic review of narrow-band imaging for | no | no | no | no | 0 | no | no | no | no | 0 | no | no | no | no | 0 | no | no | no | no | 0 |
| | the detection and differentiation of abnormalities in the esophagus and stomach (with video), Gastrointestinal Endoscopy, Volume 69, Issue 2, February 2009, Pages 307-317, ISSN 0016-5107, http://dx.doi.org/10.1016/j.gie.2008.09.048. (http://www.sciencedirect.com/science/article/pii/S0016510708026321) | | | | | | | | | | | | | | | | | | | | |
| **70** | Frank J.C. van den Broek, Johannes B. Reitsma, Wouter L. Curvers, Paul Fockens, Evelien Dekker, Systematic review of narrow-band imaging for the detection and differentiation of neoplastic and nonneoplastic lesions in the colon (with videos), Gastrointestinal Endoscopy, Volume 69, Issue 1, January 2009, Pages 124-135, ISSN 0016-5107, http://dx.doi.org/10.1016/j.gie.2008.09.040. (http://www.sciencedirect.com/science/article/pii/S001651070802628X) | no | no | no | no | 0 | no | no | no | no | 0 | no | no | no | no | 0 | no | no | no | no | 0 |
| **71** | Matthijs C Brouwer, Jan de Gans, Sebastiaan GB Heckenberg, Aeilko H Zwinderman, Tom van der Poll, Diederik van de Beek, Host genetic susceptibility to pneumococcal and meningococcal disease: a systematic review and meta-analysis, The Lancet Infectious Diseases, Volume 9, Issue 1, January 2009, Pages 31-44, ISSN 1473-3099, http://dx.doi.org/10.1016/S1473-3099(08)70261-5. (http://www.sciencedirect.com/science/article/pii/S1473309908702615) | no | no | no | yes | 1 | no | no | no | no | 0 | no | no | no | no | 0 | no | no | no | no | 0 |
| **72** | Tobias Hölscher, Søren M. Bentzen, Michael Baumann, Influence of connective tissue diseases on the expression of radiation side effects: A systematic review, Radiotherapy and Oncology, Volume 78, Issue 2, February 2006, Pages 123-130, ISSN 0167-8140, http://dx.doi.org/10.1016/j.radonc.2005.12.013. (http://www.sciencedirect.com/science/article/pii/S0167814006000065) Keywords: Radiotherapy; Radiation injuries; Radiation morbidity; Connective tissue diseases; Review; Co-morbidity | no | no | no | no | 0 | no | no | no | no | 0 | no | no | no | no | 0 | no | no | no | no | 0 |
| **73** | Ismail Celik, Lisa Gallicchio, Kristina Boyd, Tram K. Lam, Genevieve Matanoski, Xuguang Tao, Meredith Shiels, Edward Hammond, Liwei Chen, Karen A. Robinson, Laura E. Caulfield, James G. Herman, Eliseo Guallar, Anthony J. Alberg, Arsenic in drinking water and lung cancer: A systematic review, Environmental Research, Volume 108, Issue 1, September 2008, Pages 48-55, ISSN 0013-9351, http://dx.doi.org/10.1016/j.envres.2008.04.001. (http://www.sciencedirect.com/science/article/pii/S0013935108000868) Keywords: Arsenic; Drinking water; Lung cancer; Mortality; Incidence; Epidemiology; Systematic review; Taiwan | no | no | no | no | 0 | no | no | no | no | 0 | no | no | no | no | 0 | no | no | no | no | 0 |

| | | | | | | | | | | | | | | | | | | | | |
|---|---|---|---|---|---|---|---|---|---|---|---|---|---|---|---|---|---|---|---|---|
| **74** | B.S. Niël-Weise, T. Stijnen, P.J. van den Broek, Anti-infective-treated central venous catheters for total parenteral nutrition or chemotherapy: a systematic review, Journal of Hospital Infection, Volume 69, Issue 2, June 2008, Pages 114-123, ISSN 0195-6701, http://dx.doi.org/10.1016/j.jhin.2008.02.020. (http://www.sciencedirect.com/science/article/pii/S0195670108000923) Keywords: Catheter-related bloodstream infection; Central venous catheter; Antibiotics; Antiseptics; Systematic review | no | no | no | no | 0 | no | no | no | no | 0 | no | no | no | no | 0 | no | no | no | no | 0 |
| **75** | M.M.G. Leeflang, J.S. Cnossen, J.A.M. van der Post, B.W.J. Mol, K.S. Khan, G. ter Riet, Accuracy of fibronectin tests for the prediction of pre-eclampsia: a systematic review, European Journal of Obstetrics & Gynecology and Reproductive Biology, Volume 133, Issue 1, July 2007, Pages 12-19, ISSN 0301-2115, http://dx.doi.org/10.1016/j.ejogrb.2007.01.003. (http://www.sciencedirect.com/science/article/pii/S0301211507000358) Keywords: Pre-eclampsia; Prediction; Sensitivity and specificity; Fibronectin; Fetal fibronectin | no | no | no | no | 0 | no | no | no | no | 0 | no | no | no | no | 0 | no | no | no | no | 0 |
| **76** | J. Wind, S.M. Lagarde, F.J.W. ten Kate, D.T. Ubbink, W.A. Bemelman, J.J.B. van Lanschot, A systematic review on the significance of extracapsular lymph node involvement in gastrointestinal malignancies, European Journal of Surgical Oncology (EJSO), Volume 33, Issue 4, May 2007, Pages 401-408, ISSN 0748-7983, http://dx.doi.org/10.1016/j.ejso.2006.11.001. (http://www.sciencedirect.com/science/article/pii/S0748798306004525) Keywords: Lymph node metastasis; Extracapsular; Perilymphatic; Gastrointestinal malignancies; Prognosis | no | no | no | no | 0 | no | no | no | no | 0 | no | no | no | no | 0 | no | no | no | no | 0 |
| **77** | Richard M. Rosenfeld, Michael Singer, Jared M. Wasserman, Sandra S. Stinnett, Systematic review of topical antimicrobial therapy for acute otitis externa, Otolaryngology - Head and Neck Surgery, Volume 134, Issue 4, Supplement, April 2006, Pages S24-S48, ISSN 0194-5998, http://dx.doi.org/10.1016/j.otohns.2006.02.013. (http://www.sciencedirect.com/science/article/pii/S0194599806001653) | no | no | no | no | 0 | no | no | no | no | 0 | no | no | no | no | 0 | no | no | no | no | 0 |
| **78** | 2: Valderrabano P, Klippenstein DL, Tourtelot JB, Ma Z, Thompson ZJ, Lilienfeld HS, McIver B. The new ATA sonographic patterns for thyroid nodules perform well in medullary thyroid carcinoma: Institutional experience, systematic review and meta-analysis. Thyroid. 2016 Jun 7. [Epub ahead of print] PubMed PMID: | no | no | no | no | 0 | no | no | no | no | 0 | no | no | no | no | 0 | no | no | no | no | 0 |
| | 27267210. | | | | | | | | | | | | | | | | | | | | |
| **79** | 5: Goldstein BA, Navar AM, Pencina MJ, Ioannidis JP. Opportunities and challenges in developing risk prediction models with electronic health records data: a systematic review. J Am Med Inform Assoc. 2016 May 17. pii: ocw042. doi: 10.1093/jamia/ocw042. [Epub ahead of print] Review. PubMed PMID: 27189013. | yes | yes | no | no | 2 | yes | yes | yes | no | 3 | no | no | no | no | 0 | no | no | no | no | 0 |
| **80** | 6: Hege I, Kononowicz AA, Tolks D, Edelbring S, Kuehlmeyer K. A qualitative analysis of virtual patient descriptions in healthcare education based on a systematic literature review. BMC Med Educ. 2016 May 13;16(1):146. doi: 10.1186/s12909-016-0655-8. PubMed PMID: 27177766; PubMed Central PMCID: PMC4865997. | yes | yes | no | no | 2 | yes | no | no | no | 1 | no | no | no | no | 0 | no | no | no | no | 0 |
| **81** | 7: Scott JS, Sterling SA, To H, Seals SR, Jones AE. Diagnostic performance of matrix-assisted laser desorption ionisation time-of-flight mass spectrometry in blood bacterial infections: a systematic review and meta-analysis. Infect Dis (Lond). 2016 Jul;48(7):530-6. doi: 10.3109/23744235.2016.1165350. Epub 2016 Apr 27. PubMed PMID: 27118169. | no | yes | no | no | 1 | no | no | no | no | 0 | no | no | no | no | 0 | no | no | no | no | 0 |
| **82** | 8: Chang ET, Delzell E. Systematic review and meta-analysis of glyphosate exposure and risk of lymphohematopoietic cancers. J Environ Sci Health B. 2016 Jun 2;51(6):402-34. doi: 10.1080/03601234.2016.1142748. Epub 2016 Mar 25. PubMed PMID: 27015139; PubMed Central PMCID: PMC4866614. | no | no | no | no | 0 | no | no | no | no | 0 | no | no | no | no | 0 | no | no | no | no | 0 |
| **83** | 9: Leeflang MM, Ang CW, Berkhout J, Bijlmer HA, Van Bortel W, Brandenburg AH, Van Burgel ND, Van Dam AP, Dessau RB, Fingerle V, Hovius JW, Jaulhac B, Meijer B, Van Pelt W, Schellekens JF, Spijker R, Stelma FF, Stanek G, Verduyn-Lunel F, Zeller H, Sprong H. The diagnostic accuracy of serological tests for Lyme borreliosis in Europe: a systematic review and meta-analysis. BMC Infect Dis. 2016 Mar 25;16(1):140. doi: 10.1186/s12879-016-1468-4. PubMed PMID: 27013465; PubMed Central PMCID: PMC4807538. | no | no | no | no | 0 | no | no | no | no | 0 | no | no | no | no | 0 | no | no | no | no | 0 |
| **84** | 10: Ng KL, Morais C, Bernard A, Saunders N, Samaratunga H, Gobe G, Wood S. A systematic review and meta-analysis of immunohistochemical biomarkers that differentiate chromophobe renal cell carcinoma from renal oncocytoma. J Clin Pathol. 2016 Mar 7. pii: jclinpath-2015-203585. doi: 10.1136/jclinpath-2015-203585. [Epub ahead of print] Review. PubMed PMID: 26951082. | no | no | no | no | 0 | no | no | no | no | 0 | no | no | no | no | 0 | no | no | no | no | 0 |
| **85** | 12: de Groof EJ, Sahami S, Lucas C, Ponsioen CY, Bemelman WA, Buskens CJ. | no | no | no | no | 0 | no | no | no | no | 0 | no | no | no | no | 0 | no | no | no | no | 0 |
| | Treatment of perianal fistulas in Crohn's disease: a systematic review and meta-analysis comparing seton drainage and anti-TNF treatment. Colorectal Dis. 2016 Feb 27. doi: 10.1111/codi.13311. [Epub ahead of print] Review. PubMed PMID: 26921847. | | | | | | | | | | | | | | | | | | | | |
| **86** | 13: Grooten IJ, Vinke ME, Roseboom TJ, Painter RC. A Systematic Review and Meta-Analysis of the Utility of Corticosteroids in the Treatment of Hyperemesis Gravidarum. Nutr Metab Insights. 2016 Feb 4;8(Suppl 1):23-32. doi: 10.4137/NMI.S29532. eCollection 2015. Review. PubMed PMID: 26877629; PubMed Central PMCID: PMC4745642. | no | no | no | no | 0 | no | no | no | no | 0 | no | no | no | no | 0 | no | no | no | no | 0 |
| **87** | 14: Bekhuis T, Tseytlin E, Mitchell KJ. A Prototype for a Hybrid System to Support Systematic Review Teams: A Case Study of Organ Transplantation. Proceedings (IEEE Int Conf Bioinformatics Biomed). 2015 Nov;2015:940-947. PubMed PMID: 26855824; PubMed Central PMCID: PMC4742277. | no | no | no | no | 0 | no | no | no | no | 0 | no | no | no | no | 0 | no | no | no | no | 0 |
| **88** | 15: Machicado C, Marcos LA. Carcinogenesis associated with parasites other than Schistosoma, Opisthorchis and Clonorchis: A systematic review. Int J Cancer. 2016 Jun 15;138(12):2915-21. doi: 10.1002/ijc.30028. Epub 2016 Feb 19. PubMed PMID: 26840624. | no | yes | no | no | 1 | no | no | no | no | 0 | no | no | no | no | 0 | no | no | no | no | 0 |
| **89** | 16: Teroganova N, Girshkin L, Suter CM, Green MJ. DNA methylation in peripheral tissue of schizophrenia and bipolar disorder: a systematic review. BMC Genet. 2016 Jan 25;17:27. doi: 10.1186/s12863-016-0332-2. PubMed PMID: 26809779; PubMed Central PMCID: PMC4727379. | no | yes | no | yes | 2 | no | no | no | yes | 1 | no | no | no | no | 0 | no | no | no | no | 0 |
| **90** | 17: Kim K, Omori R, Ueno K, Iida S, Ito K. Host-Specific and Segment-Specific Evolutionary Dynamics of Avian and Human Influenza A Viruses: A Systematic Review. PLoS One. 2016 Jan 13;11(1):e0147021. doi: 10.1371/journal.pone.0147021. eCollection 2016. PubMed PMID: 26760775; PubMed Central PMCID: PMC4720117. | no | yes | no | no | 1 | no | no | no | no | 0 | no | no | no | no | 0 | no | no | no | no | 0 |
| **91** | 18: Cherpanath TG, Hirsch A, Geerts BF, Lagrand WK, Leeflang MM, Schultz MJ, Groeneveld AB. Predicting Fluid Responsiveness by Passive Leg Raising: A Systematic Review and Meta-Analysis of 23 Clinical Trials. Crit Care Med. 2016 May;44(5):981-91. doi: 10.1097/CCM.0000000000001556. PubMed PMID: 26741579. | no | no | no | no | 0 | no | no | no | no | 0 | no | no | no | no | 0 | no | no | no | no | 0 |
| **92** | 19: Luzak A, Schnell-Inderst P, Bühn S, Mayer-Zitarosa A, Siebert U. Clinical effectiveness of cancer screening biomarker tests offered as self-pay health service: a systematic review. Eur J Public Health. 2016 Jun;26(3):498-505. doi: 10.1093/eurpub/ckv227. Epub 2016 Jan 4. PubMed PMID: 26733629. | no | no | no | no | 0 | no | no | no | no | 0 | no | no | no | no | 0 | no | no | no | no | 0 |
| **93** | 21: Lamm SH, Ferdosi H, Dissen EK, Li J, Ahn J. A Systematic Review and Meta-Regression Analysis of Lung Cancer Risk and Inorganic Arsenic in Drinking Water. Int J Environ Res Public Health. 2015 Dec 7;12(12):15498-515. doi: 10.3390/ijerph121214990. PubMed PMID: 26690190; PubMed Central PMCID: PMC4690926. | no | no | no | no | 0 | no | no | no | no | 0 | no | no | no | no | 0 | no | no | no | no | 0 |
| **94** | 23: Dworkin JD, McKeown A, Farrar JT, Gilron I, Hunsinger M, Kerns RD, McDermott MP, Rappaport BA, Turk DC, Dworkin RH, Gewandter JS. Deficiencies in reporting of statistical methodology in recent randomized trials of nonpharmacologic pain treatments: ACTTION systematic review. J Clin Epidemiol. 2016 Apr;72:56-65. doi: 10.1016/j.jclinepi.2015.10.019. Epub 2015 Nov 18. PubMed PMID: 26597977. | no | no | no | no | 0 | no | no | no | no | 0 | no | no | no | no | 0 | no | no | no | no | 0 |
| **95** | 25: Wolthuis AM, Bislenghi G, Fieuws S, de Buck van Overstraeten A, Boeckxstaens G, D'Hoore A. Incidence of prolonged postoperative ileus after colorectal surgery: a systematic review and meta-analysis. Colorectal Dis. 2016 Jan;18(1):O1-9. doi: 10.1111/codi.13210. Review. PubMed PMID: 26558477. | no | no | no | no | 0 | no | no | no | no | 0 | no | no | no | no | 0 | no | no | no | no | 0 |
| **96** | 27: Mayo-Wilson E, Hutfless S, Li T, Gresham G, Fusco N, Ehmsen J, Heyward J, Vedula S, Lock D, Haythornthwaite J, Payne JL, Cowley T, Tolbert E, Rosman L, Twose C, Stuart EA, Hong H, Doshi P, Suarez-Cuervo C, Singh S, Dickersin K. Integrating multiple data sources (MUDS) for meta-analysis to improve patient-centered outcomes research: a protocol for a systematic review. Syst Rev. 2015 Nov 2;4:143. doi: 10.1186/s13643-015-0134-z. PubMed PMID: 26525044; PubMed Central PMCID: PMC4630908. | yes | yes | no | no | 2 | no | no | no | no | 0 | no | no | no | no | 0 | no | no | no | no | 0 |
| **97** | 28: Sherwood MW, Melloni C, Jones WS, Washam JB, Hasselblad V, Dolor RJ. Individual Proton Pump Inhibitors and Outcomes in Patients With Coronary Artery Disease on Dual Antiplatelet Therapy: A Systematic Review. J Am Heart Assoc. 2015 Oct 29;4(11). pii: e002245. doi: 10.1161/JAHA.115.002245. PubMed PMID: 26514161; PubMed Central PMCID: PMC4845227. | no | yes | no | no | 1 | no | no | no | no | 0 | no | no | no | no | 0 | no | no | no | no | 0 |
| **98** | 29: Zittermann A, Ernst JB, Birschmann I, Dittrich M. Effect of Vitamin D or | no | no | no | no | 0 | no | no | no | no | 0 | no | no | no | no | 0 | no | no | no | no | 0 |
| | Activated Vitamin D on Circulating 1,25-Dihydroxyvitamin D Concentrations: A Systematic Review and Metaanalysis of Randomized Controlled Trials. Clin Chem. 2015 Dec;61(12):1484-94. doi: 10.1373/clinchem.2015.244913. Epub 2015 Oct 28. Review. PubMed PMID: 26510958. | | | | | | | | | | | | | | | | | | | | |
| **99** | 30: Tsuji JS, Garry MR, Perez V, Chang ET. Low-level arsenic exposure and developmental neurotoxicity in children: A systematic review and risk assessment. Toxicology. 2015 Nov 4;337:91-107. doi: 10.1016/j.tox.2015.09.002. Epub 2015 Sep 18. Review. PubMed PMID: 26388044. | no | no | no | no | 0 | no | no | no | no | 0 | no | no | no | no | 0 | no | no | no | no | 0 |
| **100** | 31: Weber MA, Kleijn MH, Langendam M, Limpens J, Heineman MJ, Roovers JP. Local Oestrogen for Pelvic Floor Disorders: A Systematic Review. PLoS One. 2015 Sep 18;10(9):e0136265. doi: 10.1371/journal.pone.0136265. eCollection 2015. Review. PubMed PMID: 26383760; PubMed Central PMCID: PMC4575150. | no | no | no | no | 0 | no | no | no | no | 0 | no | no | no | no | 0 | no | no | no | no | 0 |
| **101** | 32: Roth JM, Korevaar DA, Leeflang MM, Mens PF. Molecular malaria diagnostics: A systematic review and meta-analysis. Crit Rev Clin Lab Sci. 2016;53(2):87-105. doi: 10.3109/10408363.2015.1084991. Epub 2015 Sep 17. PubMed PMID: 26376713. | no | yes | no | no | 1 | no | no | no | no | 0 | no | no | no | no | 0 | no | no | no | no | 0 |
| **102** | 33: Kadouch DJ, Schram ME, Leeflang MM, Limpens J, Spuls PI, de Rie MA. In vivo confocal microscopy of basal cell carcinoma: a systematic review of diagnostic accuracy. J Eur Acad Dermatol Venereol. 2015 Oct;29(10):1890-7. doi: 10.1111/jdv.13224. Epub 2015 Aug 19. Review. PubMed PMID: 26290493. | no | no | no | no | 0 | no | no | no | no | 0 | no | no | no | no | 0 | no | no | no | no | 0 |
| **103** | 35: Jabar A, Barth D, Matzopoulos R, Engel ME. Is the introduction of violence and injury observatories associated with a reduction of violence in adult populations? Rationale and protocol for a systematic review. BMJ Open. 2015 Jul 21;5(7):e007073. doi: 10.1136/bmjopen-2014-007073. Review. PubMed PMID: 26198425; PubMed Central PMCID: PMC4513469. | no | no | no | no | 0 | no | no | no | no | 0 | no | no | no | no | 0 | no | no | no | no | 0 |
| **104** | 36: Charlier N, Zupancic N, Fieuws S, Denhaerynck K, Zaman B, Moons P. Serious games for improving knowledge and self-management in young people with chronic conditions: a systematic review and meta-analysis. J Am Med Inform Assoc. 2016 Jan;23(1):230-9. doi: 10.1093/jamia/ocv100. Epub 2015 Jul 17. Review. PubMed PMID: 26186934. | no | no | no | no | 0 | no | no | no | no | 0 | no | no | no | no | 0 | no | no | no | no | 0 |
| **105** | 37: Dietz SM, Tacke CE, Hutten BA, Kuijpers TW. Peripheral Endothelial | no | no | no | no | 0 | no | no | no | no | 0 | no | no | no | no | 0 | no | no | no | no | 0 |
| | (Dys)Function, Arterial Stiffness and Carotid Intima-Media Thickness in Patients after Kawasaki Disease: A Systematic Review and Meta-Analyses. PLoS One. 2015 Jul 10;10(7):e0130913. doi: 10.1371/journal.pone.0130913. eCollection 2015. Review. PubMed PMID: 26161871; PubMed Central PMCID: PMC4498761. | | | | | | | | | | | | | | | | | | | | |
| **106** | 38: Coutinho FH, Meirelles PM, Moreira AP, Paranhos RP, Dutilh BE, Thompson FL. Niche distribution and influence of environmental parameters in marine microbial communities: a systematic review. PeerJ. 2015 Jun 16;3:e1008. doi: 10.7717/peerj.1008. eCollection 2015. PubMed PMID: 26157601; PubMed Central PMCID: PMC4476133. | no | yes | no | no | 1 | no | no | no | no | 0 | no | no | no | no | 0 | no | no | no | no | 0 |
| **107** | 39: Foulds JA, Adamson SJ, Boden JM, Williman JA, Mulder RT. Depression in patients with alcohol use disorders: Systematic review and meta-analysis of outcomes for independent and substance-induced disorders. J Affect Disord. 2015 Oct 1;185:47-59. doi: 10.1016/j.jad.2015.06.024. Epub 2015 Jun 23. Review. PubMed PMID: 26143404. | no | no | no | no | 0 | no | no | no | no | 0 | no | no | no | no | 0 | no | no | no | no | 0 |
| **108** | 40: Champredon D, Bellan SE, Delva W, Hunt S, Shi CF, Smieja M, Dushoff J. The effect of sexually transmitted co-infections on HIV viral load amongst individuals on antiretroviral therapy: a systematic review and meta-analysis. BMC Infect Dis. 2015 Jun 30;15:249. doi: 10.1186/s12879-015-0961-5. Review. PubMed PMID: 26123030; PubMed Central PMCID: PMC4486691. | no | no | no | no | 0 | no | no | no | no | 0 | no | no | no | no | 0 | no | no | no | no | 0 |
| **109** | 41: Salnikova LE, Kolobkov DS. Germline and somatic genetic predictors of pathological response in neoadjuvant settings of rectal and esophageal cancers: systematic review and meta-analysis. Pharmacogenomics J. 2016 Jun;16(3):249-65. doi: 10.1038/tpj.2015.46. Epub 2015 Jun 30. PubMed PMID: 26122021. | no | yes | no | yes | 2 | no | no | no | yes | 1 | no | no | no | no | 0 | no | no | no | no | 0 |
| **110** | 42: Nunes LA, Mussavira S, Bindhu OS. Clinical and diagnostic utility of saliva as a non-invasive diagnostic fluid: a systematic review. Biochem Med (Zagreb). 2015 Jun 5;25(2):177-92. doi: 10.11613/BM.2015.018. eCollection 2015. Review. PubMed PMID: 26110030; PubMed Central PMCID: PMC4470107. | no | no | no | no | 0 | no | no | no | no | 0 | no | no | no | no | 0 | no | no | no | no | 0 |
| **111** | 43: Law LS, Tan M, Bai Y, Miller TE, Li YJ, Gan TJ. Paravertebral Block for Inguinal Herniorrhaphy: A Systematic Review and Meta-Analysis of Randomized Controlled Trials. Anesth Analg. 2015 Aug;121(2):556-69. doi: 10.1213/ANE.0000000000000835. Review. PubMed PMID: 26086619. | no | no | no | no | 0 | no | no | no | no | 0 | no | no | no | no | 0 | no | no | no | no | 0 |
| **112** | 44: Ramirez-Busby SM, Valafar F. Systematic review of mutations in pyrazinamidase associated with pyrazinamide resistance in Mycobacterium tuberculosis clinical isolates. Antimicrob Agents Chemother. 2015 Sep;59(9):5267-77. doi: 10.1128/AAC.00204-15. Epub 2015 Jun 15. Review. PubMed PMID: 26077261; PubMed Central PMCID: PMC4538510. | yes | yes | no | yes | 3 | yes | yes | yes | yes | 4 | yes | no | no | yes | 2 | yes | yes | yes | yes | 4 |
| **113** | 46: Ali-Khan SE, Black L, Palmour N, Hallett MT, Avard D. SOCIO-ETHICAL ISSUES IN PERSONALIZED MEDICINE: A SYSTEMATIC REVIEW OF ENGLISH LANGUAGE HEALTH TECHNOLOGY ASSESSMENTS OF GENE EXPRESSION PROFILING TESTS FOR BREAST CANCER PROGNOSIS. Int J Technol Assess Health Care. 2015 Jan;31(1-2):36-50. doi: 10.1017/S0266462315000082. Epub 2015 May 20. PubMed PMID: 25991501. | yes | yes | no | no | 2 | no | yes | no | no | 1 | no | no | no | no | 0 | no | no | no | no | 0 |
| **114** | 48: Punt S, Langenhoff JM, Putter H, Fleuren GJ, Gorter A, Jordanova ES. The correlations between IL-17 vs. Th17 cells and cancer patient survival: a systematic review. Oncoimmunology. 2015 Mar 6;4(2):e984547. eCollection 2015 Feb. Review. PubMed PMID: 25949881; PubMed Central PMCID: PMC4404813. | no | yes | no | no | 1 | no | no | no | no | 0 | no | no | no | no | 0 | no | no | no | no | 0 |
| **115** | 49: van der Voort P, Pijls BG, Nieuwenhuijse MJ, Jasper J, Fiocco M, Plevier JW, Middeldorp S, Valstar ER, Nelissen RG. Early subsidence of shape-closed hip arthroplasty stems is associated with late revision. A systematic review and meta-analysis of 24 RSA studies and 56 survival studies. Acta Orthop. 2015;86(5):575-85. doi: 10.3109/17453674.2015.1043832. Review. PubMed PMID: 25909455; PubMed Central PMCID: PMC4564780. | no | no | no | no | 0 | no | no | no | no | 0 | no | no | no | no | 0 | no | no | no | no | 0 |
| **116** | 50: Singla N, Hunsinger M, Chang PD, McDermott MP, Chowdhry AK, Desjardins PJ, Turk DC, Dworkin RH. Assay sensitivity of pain intensity versus pain relief in acute pain clinical trials: ACTTION systematic review and meta-analysis. J Pain. 2015 Aug;16(8):683-91. doi: 10.1016/j.jpain.2015.03.015. Epub 2015 Apr 16. Review. PubMed PMID: 25892656. | no | no | no | no | 0 | no | no | no | no | 0 | no | no | no | no | 0 | no | no | no | no | 0 |
| **117** | 51: Smeulers M, Verweij L, Maaskant JM, de Boer M, Krediet CT, Nieveen van Dijkum EJ, Vermeulen H. Quality indicators for safe medication preparation and administration: a systematic review. PLoS One. 2015 Apr 17;10(4):e0122695. doi: 10.1371/journal.pone.0122695. eCollection 2015. Review. PubMed PMID: 25884623; PubMed Central PMCID: PMC4401721. | no | no | no | no | 0 | no | no | no | no | 0 | no | no | no | no | 0 | no | no | no | no | 0 |
| **118** | 52: Sejersen MH, Frost P, Hansen TB, Deutch SR, Svendsen SW. Proteomics | yes | yes | yes | yes | 4 | yes | yes | yes | yes | 4 | yes | yes | yes | no | 3 | yes | yes | yes | no | 3 |
| | perspectives in rotator cuff research: a systematic review of gene expression and protein composition in human tendinopathy. PLoS One. 2015 Apr 16;10(4):e0119974. doi: 10.1371/journal.pone.0119974. eCollection 2015. Review. PubMed PMID: 25879758; PubMed Central PMCID: PMC4400011. | | | | | | | | | | | | | | | | | | | | |
| **119** | 53: Seijmonsbergen-Schermers AE, Sahami S, Lucas C, Jonge Ad. Nonsuturing or Skin Adhesives versus Suturing of the Perineal Skin After Childbirth: A Systematic Review. Birth. 2015 Jun;42(2):100-15. doi: 10.1111/birt.12166. Epub 2015 Apr 11. Review. PubMed PMID: 25864727. | no | no | no | no | 0 | no | no | no | no | 0 | no | no | no | no | 0 | no | no | no | no | 0 |
| **120** | 54: Thijssen VL, Heusschen R, Caers J, Griffioen AW. Galectin expression in cancer diagnosis and prognosis: A systematic review. Biochim Biophys Acta. 2015 Apr;1855(2):235-47. doi: 10.1016/j.bbcan.2015.03.003. Epub 2015 Mar 25. Review. PubMed PMID: 25819524. | yes | yes | no | no | 2 | no | no | no | yes | 1 | no | no | no | no | 0 | no | no | no | no | 0 |
| **121** | 56: Kaufman JS, Dolman L, Rushani D, Cooper RS. The contribution of genomic research to explaining racial disparities in cardiovascular disease: a systematic review. Am J Epidemiol. 2015 Apr 1;181(7):464-72. doi: 10.1093/aje/kwu319. Epub 2015 Mar 1. Review. PubMed PMID: 25731887. | yes | yes | no | yes | 3 | yes | yes | yes | yes | 4 | yes | yes | no | yes | 3 | yes | yes | no | no | 2 |
| **122** | 57: Smith SM, Hunsinger M, McKeown A, Parkhurst M, Allen R, Kopko S, Lu Y, Wilson HD, Burke LB, Desjardins P, McDermott MP, Rappaport BA, Turk DC, Dworkin RH. Quality of pain intensity assessment reporting: ACTTION systematic review and recommendations. J Pain. 2015 Apr;16(4):299-305. doi: 10.1016/j.jpain.2015.01.004. Epub 2015 Jan 28. Review. PubMed PMID: 25637296. | no | no | no | no | 0 | no | no | no | no | 0 | no | no | no | no | 0 | no | no | no | no | 0 |
| **123** | 58: Kulkarni M, Gokulakrishnan G, Price J, Fernandes CJ, Leeflang M, Pammi M. Diagnosing significant PDA using natriuretic peptides in preterm neonates: a systematic review. Pediatrics. 2015 Feb;135(2):e510-25. doi: 10.1542/peds.2014-1995. Epub 2015 Jan 19. Review. PubMed PMID: 25601976. | yes | yes | no | no | 2 | no | no | no | no | 0 | no | no | no | no | 0 | no | no | no | no | 0 |
| **124** | 59: O'Mara-Eves A, Thomas J, McNaught J, Miwa M, Ananiadou S. Using text mining for study identification in systematic reviews: a systematic review of current approaches. Syst Rev. 2015 Jan 14;4:5. doi: 10.1186/2046-4053-4-5. Review. Erratum in: Syst Rev. 2015;4:59. PubMed PMID: 25588314; PubMed Central PMCID: PMC4320539. | yes | yes | no | yes | 3 | yes | yes | yes | yes | 4 | yes | yes | yes | yes | 4 | no | no | no | no | 0 |
| **125** | 60: Hill LK, Hu DD, Koenig J, Sollers JJ 3rd, Kapuku G, Wang X, Snieder H, Thayer JF. Ethnic differences in resting heart rate variability: a systematic review and meta-analysis. Psychosom Med. 2015 Jan;77(1):16-25. doi: 10.1097/PSY.0000000000000133. Review. PubMed PMID: 25551201; PubMed Central PMCID: PMC4293235. | no | no | no | no | 0 | no | no | no | no | 0 | no | no | no | no | 0 | no | no | no | no | 0 |
| **126** | 61: Keurentjes JC, Pijls BG, Van Tol FR, Mentink JF, Mes SD, Schoones JW, Fiocco M, Sedrakyan A, Nelissen RG. Which implant should we use for primary total hip replacement? A systematic review and meta-analysis. J Bone Joint Surg Am. 2014 Dec 17;96 Suppl 1:79-97. doi: 10.2106/JBJS.N.00397. Review. PubMed PMID: 25520423. | no | no | no | no | 0 | no | no | no | no | 0 | no | no | no | no | 0 | no | no | no | no | 0 |
| **127** | 62: McKeown A, Gewandter JS, McDermott MP, Pawlowski JR, Poli JJ, Rothstein D, Farrar JT, Gilron I, Katz NP, Lin AH, Rappaport BA, Rowbotham MC, Turk DC, Dworkin RH, Smith SM. Reporting of sample size calculations in analgesic clinical trials: ACTTION systematic review. J Pain. 2015 Mar;16(3):199-206.e1-7. doi: 10.1016/j.jpain.2014.11.010. Epub 2014 Dec 4. Review. PubMed PMID: 25481494. | no | no | no | no | 0 | no | no | no | no | 0 | no | no | no | no | 0 | no | no | no | no | 0 |
| **128** | 63: Uronis HE, Ekström MP, Currow DC, McCrory DC, Samsa GP, Abernethy AP. Oxygen for relief of dyspnoea in people with chronic obstructive pulmonary disease who would not qualify for home oxygen: a systematic review and meta-analysis. Thorax. 2015 May;70(5):492-4. doi: 10.1136/thoraxjnl-2014-205720. Epub 2014 Dec 3. Review. PubMed PMID: 25472664. | no | no | no | no | 0 | no | no | no | no | 0 | no | no | no | no | 0 | no | no | no | no | 0 |
| **129** | 64: Gewandter JS, McKeown A, McDermott MP, Dworkin JD, Smith SM, Gross RA, Hunsinger M, Lin AH, Rappaport BA, Rice AS, Rowbotham MC, Williams MR, Turk DC, Dworkin RH. Data interpretation in analgesic clinical trials with statistically nonsignificant primary analyses: an ACTTION systematic review. J Pain. 2015 Jan;16(1):3-10. doi: 10.1016/j.jpain.2014.10.003. Epub 2014 Oct 23. Review. PubMed PMID: 25451621. | no | yes | no | no | 1 | no | no | no | no | 0 | no | no | no | no | 0 | no | no | no | no | 0 |
| **130** | 65: Sapkota Y. Germline DNA variations in breast cancer predisposition and prognosis: a systematic review of the literature. Cytogenet Genome Res. 2014;144(2):77-91. doi: 10.1159/000369045. Epub 2014 Nov 15. Review. PubMed PMID: 25401968. | no | no | no | yes | 1 | no | no | no | no | 0 | no | no | no | no | 0 | no | no | no | no | 0 |
| **131** | 67: Khan GH, Galazis N, Docheva N, Layfield R, Atiomo W. Overlap of | yes | yes | no | no | 2 | yes | no | no | yes | 2 | no | no | no | no | 0 | no | no | no | no | 0 |
| | proteomics biomarkers between women with pre-eclampsia and PCOS: a systematic review and biomarker database integration. Hum Reprod. 2015 Jan;30(1):133-48. doi: 10.1093/humrep/deu268. Epub 2014 Oct 28. Review. PubMed PMID: 25351721; PubMed Central PMCID: PMC4262466. | | | | | | | | | | | | | | | | | | | | |
| **132** | 68: Daoud FC. Systematic literature review update of the PROUD trial: potential usefulness of a collaborative database. Surg Infect (Larchmt). 2014 Dec;15(6):857-8. doi: 10.1089/sur.2014.129. Review. PubMed PMID: 25317784; PubMed Central PMCID: PMC4268572. | yes | yes | no | yes | 3 | no | no | no | yes | 1 | no | no | no | no | 0 | no | no | no | no | 0 |
| **133** | 69: Gewandter JS, McDermott MP, McKeown A, Smith SM, Pawlowski JR, Poli JJ, Rothstein D, Williams MR, Bujanover S, Farrar JT, Gilron I, Katz NP, Rowbotham MC, Turk DC, Dworkin RH. Reporting of intention-to-treat analyses in recent analgesic clinical trials: ACTTION systematic review and recommendations. Pain. 2014 Dec;155(12):2714-9. doi: 10.1016/j.pain.2014.09.039. Epub 2014 Oct 2. Review. PubMed PMID: 25284072. | no | no | no | no | 0 | no | no | no | no | 0 | no | no | no | no | 0 | no | no | no | no | 0 |
| **134** | 70: Shabani M, Bezuidenhout L, Borry P. Attitudes of research participants and the general public towards genomic data sharing: a systematic literature review. Expert Rev Mol Diagn. 2014 Nov;14(8):1053-65. doi: 10.1586/14737159.2014.961917. Epub 2014 Sep 26. Review. PubMed PMID: 25260013. | yes | yes | no | yes | 3 | yes | yes | yes | yes | 4 | yes | yes | no | no | 2 | no | no | no | yes | 1 |
| **135** | 72: Huynh J, Xiong G, Bentley-Lewis R. A systematic review of metabolite profiling in gestational diabetes mellitus. Diabetologia. 2014 Dec;57(12):2453-64. doi: 10.1007/s00125-014-3371-0. Epub 2014 Sep 6. Review. PubMed PMID: 25193282; PubMed Central PMCID: PMC4221524. | yes | no | no | no | 1 | no | no | no | no | 0 | no | no | no | no | 0 | no | no | no | no | 0 |
| **136** | 73: Hudson CO, Northington GM, Lyles RH, Karp DR. Outcomes of robotic sacrocolpopexy: a systematic review and meta-analysis. Female Pelvic Med Reconstr Surg. 2014 Sep-Oct;20(5):252-60. doi: 10.1097/SPV.0000000000000070. Review. PubMed PMID: 25181374; PubMed Central PMCID: PMC4374352. | yes | yes | no | no | 2 | no | no | no | no | 0 | no | no | no | no | 0 | no | no | no | no | 0 |
| **137** | 74: Coelho T, Andreoletti G, Ashton JJ, Pengelly RJ, Gao Y, RamaKrishnan A, Batra A, Beattie RM, Williams AP, Ennis S. Immuno-genomic profiling of patients with inflammatory bowel disease: a systematic review of genetic and functional in vivo studies of implicated genes. Inflamm Bowel Dis. 2014 Oct;20(10):1813-9. doi: 10.1097/MIB.0000000000000174. Review. PubMed PMID: 25171511. | yes | yes | yes | yes | 4 | yes | yes | yes | yes | 4 | no | no | no | yes | 1 | no | no | no | no | 0 |
| **138** | 75: Hunsinger M, Smith SM, Rothstein D, McKeown A, Parkhurst M, Hertz S, Katz NP, Lin AH, McDermott MP, Rappaport BA, Turk DC, Dworkin RH. Adverse event reporting in nonpharmacologic, noninterventional pain clinical trials: ACTTION systematic review. Pain. 2014 Nov;155(11):2253-62. doi: 10.1016/j.pain.2014.08.004. Epub 2014 Aug 12. Review. PubMed PMID: 25123543. | no | no | no | no | 0 | no | no | no | no | 0 | no | no | no | no | 0 | no | no | no | no | 0 |
| **139** | 78: Vitek WS, Shayne M, Hoeger K, Han Y, Messing S, Fung C. Gonadotropin-releasing hormone agonists for the preservation of ovarian function among women with breast cancer who did not use tamoxifen after chemotherapy: a systematic review and meta-analysis. Fertil Steril. 2014 Sep;102(3):808-815.e1. doi: 10.1016/j.fertnstert.2014.06.003. Epub 2014 Jul 17. Review. PubMed PMID: 25044080. | no | no | no | no | 0 | no | no | no | no | 0 | no | no | no | no | 0 | no | no | no | no | 0 |
| **140** | 79: Gewandter JS, McDermott MP, McKeown A, Smith SM, Williams MR, Hunsinger M, Farrar J, Turk DC, Dworkin RH. Reporting of missing data and methods used to accommodate them in recent analgesic clinical trials: ACTTION systematic review and recommendations. Pain. 2014 Sep;155(9):1871-7. doi: 10.1016/j.pain.2014.06.018. Epub 2014 Jun 30. Review. PubMed PMID: 24993384. | yes | no | no | no | 1 | no | no | no | no | 0 | no | no | no | no | 0 | no | no | no | no | 0 |
| **141** | 80: Tao Y, Wang Y, Rogers JT, Wang F. Perturbed iron distribution in Alzheimer's disease serum, cerebrospinal fluid, and selected brain regions: a systematic review and meta-analysis. J Alzheimers Dis. 2014;42(2):679-90. doi: 10.3233/JAD-140396. Review. PubMed PMID: 24916541. | no | no | no | no | 0 | no | no | no | no | 0 | no | no | no | no | 0 | no | no | no | no | 0 |
| **142** | 82: Aslam S, Vaida F, Ritter M, Mehta RL. Systematic review and meta-analysis on management of hemodialysis catheter-related bacteremia. J Am Soc Nephrol. 2014 Dec;25(12):2927-41. doi: 10.1681/ASN.2013091009. Epub 2014 May 22. Review. PubMed PMID: 24854263; PubMed Central PMCID: PMC4243345. | no | no | no | no | 0 | no | no | no | no | 0 | no | no | no | no | 0 | no | no | no | no | 0 |
| **143** | 83: Bălănescu P, Lădaru A, Bălănescu E, Băicuş C, Dan GA. Systemic sclerosis biomarkers discovered using mass-spectrometry-based proteomics: a systematic review. Biomarkers. 2014 Aug;19(5):345-55. doi: 10.3109/1354750X.2014.920046. Epub 2014 May 16. Review. PubMed PMID: 24831309. | yes | yes | no | yes | 3 | yes | yes | yes | yes | 4 | no | no | no | no | 0 | no | no | no | no | 0 |
| **144** | 84: de Ruiter CM, van der Veer C, Leeflang MM, Deborggraeve S, Lucas C, Adams ER. Molecular tools for diagnosis of visceral leishmaniasis: systematic review and meta-analysis of diagnostic test accuracy. J Clin Microbiol. 2014 Sep;52(9):3147- | yes | yes | yes | no | 3 | no | no | no | no | 0 | no | no | no | no | 0 | no | no | no | no | 0 |
| | 55. doi: 10.1128/JCM.00372-14. Epub 2014 May 14. Review. PubMed PMID: 24829226; PubMed Central PMCID: PMC4313130. | | | | | | | | | | | | | | | | | | | | |
| **145** | 85: Li X, Shen L, Tan H. Polymorphisms and plasma level of transforming growth factor-Beta 1 and risk for preeclampsia: a systematic review. PLoS One. 2014 May 13;9(5):e97230. doi: 10.1371/journal.pone.0097230. eCollection 2014. Review. PubMed PMID: 24823830; PubMed Central PMCID: PMC4019528. | no | no | no | no | 0 | no | no | no | no | 0 | no | no | no | no | 0 | no | no | no | no | 0 |
| **146** | 86: Olesen TB, Munk C, Christensen J, Andersen KK, Kjaer SK. Human papillomavirus prevalence among men in sub-Saharan Africa: a systematic review and meta-analysis. Sex Transm Infect. 2014 Sep;90(6):455-62. doi: 10.1136/sextrans-2013-051456. Epub 2014 May 7. Review. PubMed PMID: 24812407. | no | no | no | no | 0 | no | no | no | no | 0 | no | no | no | no | 0 | no | no | no | no | 0 |
| **147** | 87: Edefonti V, Rosato V, Parpinel M, Nebbia G, Fiorica L, Fossali E, Ferraroni M, Decarli A, Agostoni C. The effect of breakfast composition and energy contribution on cognitive and academic performance: a systematic review. Am J Clin Nutr. 2014 Aug;100(2):626-56. doi: 10.3945/ajcn.114.083683. Epub 2014 May 7. Review. PubMed PMID: 24808492. | no | no | no | no | 0 | no | no | no | no | 0 | no | no | no | no | 0 | no | no | no | no | 0 |
| **148** | 88: Rostamian S, Mahinrad S, Stijnen T, Sabayan B, de Craen AJ. Cognitive impairment and risk of stroke: a systematic review and meta-analysis of prospective cohort studies. Stroke. 2014 May;45(5):1342-8. doi: 10.1161/STROKEAHA.114.004658. Epub 2014 Mar 27. Review. PubMed PMID: 24676778. | no | no | no | no | 0 | no | no | no | no | 0 | no | no | no | no | 0 | no | no | no | no | 0 |
| **149** | 92: Nglazi MD, Bekker LG, Wood R, Shey MS, Uthman OA, Wiysonge CS. The impact of mass media interventions on tuberculosis awareness, health-seeking behaviour and health service utilisation: a systematic review protocol. BMJ Open. 2014 Jan 14;4(1):e004302. doi: 10.1136/bmjopen-2013-004302. Review. PubMed PMID: 24430882; PubMed Central PMCID: PMC3902379. | no | no | no | no | 0 | no | no | no | no | 0 | no | no | no | no | 0 | no | no | no | no | 0 |
| **150** | 93: Mocchegiani E, Costarelli L, Giacconi R, Malavolta M, Basso A, Piacenza F, Ostan R, Cevenini E, Gonos ES, Monti D. Micronutrient-gene interactions related to inflammatory/immune response and antioxidant activity in ageing and inflammation. A systematic review. Mech Ageing Dev. 2014 Mar-Apr;136-137:29-49. doi: 10.1016/j.mad.2013.12.007. Epub 2014 Jan 2. Review. PubMed PMID: 24388876. | yes | yes | no | no | 2 | no | no | no | yes | 1 | no | no | no | no | 0 | no | no | no | no | 0 |
| **151** | 94: Nsoesie EO, Brownstein JS, Ramakrishnan N, Marathe MV. A systematic review of studies on forecasting the dynamics of influenza outbreaks. Influenza Other Respir Viruses. 2014 May;8(3):309-16. doi: 10.1111/irv.12226. Epub 2013 Dec 23. Review. PubMed PMID: 24373466; PubMed Central PMCID: PMC4181479. | yes | yes | no | no | 2 | no | yes | no | no | 1 | no | no | no | no | 0 | no | no | no | no | 0 |
| **152** | 95: Korevaar DA, van Enst WA, Spijker R, Bossuyt PM, Hooft L. Reporting quality of diagnostic accuracy studies: a systematic review and meta-analysis of investigations on adherence to STARD. Evid Based Med. 2014 Apr;19(2):47-54. doi: 10.1136/eb-2013-101637. Epub 2013 Dec 24. Review. PubMed PMID: 24368333. | no | no | no | no | 0 | no | no | no | no | 0 | no | no | no | no | 0 | no | no | no | no | 0 |
| **153** | 96: Pease M, Ling C, Mack WJ, Wang K, Zada G. The role of epigenetic modification in tumorigenesis and progression of pituitary adenomas: a systematic review of the literature. PLoS One. 2013 Dec 18;8(12):e82619. doi: 10.1371/journal.pone.0082619. eCollection 2013. Review. PubMed PMID: 24367530; PubMed Central PMCID: PMC3867353. | yes | yes | no | yes | 3 | yes | yes | no | yes | 3 | no | no | no | no | 0 | no | no | no | no | 0 |
| **154** | 97: Schleidgen S, Klingler C, Bertram T, Rogowski WH, Marckmann G. What is personalized medicine: sharpening a vague term based on a systematic literature review. BMC Med Ethics. 2013 Dec 21;14:55. doi: 10.1186/1472-6939-14-55. Review. PubMed PMID: 24359531; PubMed Central PMCID: PMC3878093. | no | yes | no | no | 1 | no | no | no | no | 0 | no | no | no | no | 0 | no | no | no | no | 0 |
| **155** | 98: Hunsinger M, Smith SM, McKeown A, Parkhurst M, Gross RA, Lin AH, McDermott MP, Rappaport BA, Turk DC, Dworkin RH. Disclosure of authorship contributions in analgesic clinical trials and related publications: ACTTION systematic review and recommendations. Pain. 2014 Jun;155(6):1059-63. doi: 10.1016/j.pain.2013.12.011. Epub 2013 Dec 12. Review. PubMed PMID: 24334187. | no | no | no | no | 0 | no | no | no | no | 0 | no | no | no | no | 0 | no | no | no | no | 0 |
| **156** | 99: Oren A, Garrity GM. Then and now: a systematic review of the systematics of prokaryotes in the last 80 years. Antonie Van Leeuwenhoek. 2014 Jul;106(1):43-56. doi: 10.1007/s10482-013-0084-1. Epub 2013 Dec 5. Review. PubMed PMID: 24306768. | no | yes | no | no | 1 | no | no | no | no | 0 | no | no | no | no | 0 | no | no | no | no | 0 |
| **157** | 100: Gewandter JS, Smith SM, McKeown A, Burke LB, Hertz SH, Hunsinger M, Katz NP, Lin AH, McDermott MP, Rappaport BA, Williams MR, Turk DC, Dworkin RH. Reporting of primary analyses and multiplicity adjustment in recent | no | no | no | no | 0 | no | no | no | no | 0 | no | no | no | no | 0 | no | no | no | no | 0 |
| | analgesic clinical trials: ACTTION systematic review and recommendations. Pain. 2014 Mar;155(3):461-6. doi: 10.1016/j.pain.2013.11.009. Epub 2013 Nov 23. Review. PubMed PMID: 24275257. | | | | | | | | | | | | | | | | | | | | |
| **158** | 102: Kacerovsky M, Lenco J, Musilova I, Tambor V, Lamont R, Torloni MR, Menon R; PREBIC Biomarker Working Group 2012-2013. Proteomic biomarkers for spontaneous preterm birth: a systematic review of the literature. Reprod Sci. 2014 Mar;21(3):283-95. doi: 10.1177/1933719113503415. Epub 2013 Sep 23. Review. PubMed PMID: 24060632. | yes | yes | no | yes | 3 | yes | yes | yes | yes | 4 | no | yes | no | no | 1 | no | no | no | no | 0 |
| **159** | 103: Gierisch JM, Coeytaux RR, Urrutia RP, Havrilesky LJ, Moorman PG, Lowery WJ, Dinan M, McBroom AJ, Hasselblad V, Sanders GD, Myers ER. Oral contraceptive use and risk of breast, cervical, colorectal, and endometrial cancers: a systematic review. Cancer Epidemiol Biomarkers Prev. 2013 Nov;22(11):1931-43. doi: 10.1158/1055-9965.EPI-13-0298. Epub 2013 Sep 6. Review. PubMed PMID: 24014598. | no | no | no | no | 0 | no | no | no | no | 0 | no | no | no | no | 0 | no | no | no | no | 0 |
| **160** | 104: de Wit MC, Srebniak MI, Govaerts LC, Van Opstal D, Galjaard RJ, Go AT. Additional value of prenatal genomic array testing in fetuses with isolated structural ultrasound abnormalities and a normal karyotype: a systematic review of the literature. Ultrasound Obstet Gynecol. 2014 Feb;43(2):139-46. doi: 10.1002/uog.12575. Review. PubMed PMID: 23897843. | yes | yes | no | no | 2 | no | no | no | no | 0 | no | no | no | no | 0 | no | no | no | no | 0 |
| **161** | 105: Abbassi-Ghadi N, Kumar S, Huang J, Goldin R, Takats Z, Hanna GB. Metabolomic profiling of oesophago-gastric cancer: a systematic review. Eur J Cancer. 2013 Nov;49(17):3625-37. doi: 10.1016/j.ejca.2013.07.004. Epub 2013 Jul 26. Review. PubMed PMID: 23896378. | yes | no | no | no | 1 | no | no | no | no | 0 | no | no | no | no | 0 | no | no | no | no | 0 |
| **162** | 106: Jiang F, Zhou XY, Huang J. The value of surface enhanced laser desorption/ionization-time of flight mass spectrometry at the diagnosis of non-small cell lung cancer: a systematic review. Technol Cancer Res Treat. 2014 Apr;13(2):109-17. doi: 10.7785/tcrt.2012.500360. Epub 2013 Jul 11. Review. PubMed PMID: 23862745. | yes | no | no | no | 1 | no | no | no | no | 0 | no | no | no | no | 0 | no | no | no | no | 0 |
| **163** | 107: Gopalakrishna G, Langendam MW, Scholten RJ, Bossuyt PM, Leeflang MM. Guidelines for guideline developers: a systematic review of grading systems for medical tests. Implement Sci. 2013 Jul 10;8:78. doi: 10.1186/1748-5908-8-78. Review. PubMed PMID: 23842037; PubMed Central PMCID: PMC3716938. | yes | no | no | yes | 2 | no | no | no | no | 0 | no | no | no | no | 0 | no | no | no | no | 0 |
| **164** | 108: de Viron S, Malats N, Van der Heyden J, Van Oyen H, Brand A. Environmental and genomic factors as well as interventions influencing smoking cessation: a systematic review of reviews and a proposed working model. Public Health Genomics. 2013;16(4):159-73. doi: 10.1159/000351453. Epub 2013 Jun 21. Review. PubMed PMID: 23796797. | yes | yes | no | yes | 3 | no | no | no | no | 0 | no | no | no | no | 0 | no | no | no | no | 0 |
| **165** | 109: Tajik P, Zwinderman AH, Mol BW, Bossuyt PM. Trial designs for personalizing cancer care: a systematic review and classification. Clin Cancer Res. 2013 Sep 1;19(17):4578-88. doi: 10.1158/1078-0432.CCR-12-3722. Epub 2013 Jun 20. Review. PubMed PMID: 23788580. | yes | yes | no | no | 2 | no | no | no | no | 0 | no | no | no | no | 0 | no | no | no | no | 0 |
| **166** | 110: Wright OR. Systematic review of knowledge, confidence and education in nutritional genomics for students and professionals in nutrition and dietetics. J Hum Nutr Diet. 2014 Jun;27(3):298-307. doi: 10.1111/jhn.12132. Epub 2013 Jun 20. Review. PubMed PMID: 23781868. | yes | yes | no | no | 2 | no | yes | no | no | 1 | no | no | no | no | 0 | no | no | no | no | 0 |
| **167** | 111: Weber MT, Maki PM, McDermott MP. Cognition and mood in perimenopause: a systematic review and meta-analysis. J Steroid Biochem Mol Biol. 2014 Jul;142:90-8. doi: 10.1016/j.jsbmb.2013.06.001. Epub 2013 Jun 14. Review. PubMed PMID: 23770320; PubMed Central PMCID: PMC3830624. | no | no | no | no | 0 | no | no | no | no | 0 | no | no | no | no | 0 | no | no | no | no | 0 |
| **168** | 112: Snoeker BA, Bakker EW, Kegel CA, Lucas C. Risk factors for meniscal tears: a systematic review including meta-analysis. J Orthop Sports Phys Ther. 2013 Jun;43(6):352-67. doi: 10.2519/jospt.2013.4295. Epub 2013 Apr 29. Review. PubMed PMID: 23628788. | no | no | no | no | 0 | no | no | no | no | 0 | no | no | no | no | 0 | no | no | no | no | 0 |
| **169** | 113: Zhang AH, Sun H, Yan GL, Han Y, Wang XJ. Serum proteomics in biomedical research: a systematic review. Appl Biochem Biotechnol. 2013 Jun;170(4):774-86. doi: 10.1007/s12010-013-0238-7. Epub 2013 Apr 23. Review. PubMed PMID: 23609910. | yes | yes | no | yes | 3 | yes | yes | yes | yes | 4 | no | no | no | no | 0 | no | no | no | no | 0 |
| **170** | 114: Galazis N, Pang YL, Galazi M, Haoula Z, Layfield R, Atiomo W. Proteomic biomarkers of endometrial cancer risk in women with polycystic ovary syndrome: a systematic review and biomarker database integration. Gynecol Endocrinol. 2013 Jul;29(7):638-44. doi: 10.3109/09513590.2013.777416. Epub 2013 Mar 25. Review. PubMed PMID: 23527552. | yes | yes | yes | no | 3 | yes | yes | yes | no | 3 | no | no | no | no | 0 | no | no | no | no | 0 |
| **171** | 115: Knops AM, Legemate DA, Goossens A, Bossuyt PM, Ubbink DT. Decision | no | no | no | no | 0 | no | no | no | no | 0 | no | no | no | no | 0 | no | no | no | no | 0 |
| | aids for patients facing a surgical treatment decision: a systematic review and meta-analysis. Ann Surg. 2013 May;257(5):860-6. doi: 10.1097/SLA.0b013e3182864fd6. Review. PubMed PMID: 23470574. | | | | | | | | | | | | | | | | | | | | |
| **172** | 116: Jonnalagadda S, Petitti D. A new iterative method to reduce workload in systematic review process. Int J Comput Biol Drug Des. 2013;6(1-2):5-17. doi: 10.1504/IJCBDD.2013.052198. Epub 2013 Feb 21. PubMed PMID: 23428470; PubMed Central PMCID: PMC3787693. | yes | yes | no | no | 2 | yes | no | no | yes | 2 | no | no | no | no | 0 | no | no | no | no | 0 |
| **173** | 117: Baumgartel KL, Conley YP. The utility of breastmilk for genetic or genomic studies: a systematic review. Breastfeed Med. 2013 Jun;8(3):249-56. doi: 10.1089/bfm.2012.0054. Epub 2012 Dec 21. Review. PubMed PMID: 23259645; PubMed Central PMCID: PMC3663450. | yes | yes | no | yes | 3 | yes | yes | yes | no | 3 | no | no | no | no | 0 | no | no | no | no | 0 |
| **174** | 118: Wang X, Zhang A, Sun H, Wang P. Systems biology technologies enable personalized traditional Chinese medicine: a systematic review. Am J Chin Med. 2012;40(6):1109-22. doi: 10.1142/S0192415X12500826. Review. PubMed PMID: 23227785. | yes | yes | yes | no | 3 | no | yes | yes | no | 2 | yes | yes | yes | no | 3 | no | no | no | no | 0 |
| **175** | 119: Ma Y, Zhang P, Wang F, Qin H. Searching for consistently reported up- and down-regulated biomarkers in colorectal cancer: a systematic review of proteomic studies. Mol Biol Rep. 2012 Aug;39(8):8483-90. doi: 10.1007/s11033-012-1702-0. Epub 2012 Jun 15. Review. PubMed PMID: 22699879. | yes | yes | no | yes | 3 | yes | yes | yes | no | 3 | no | no | no | no | 0 | no | no | no | no | 0 |
| **176** | 120: Zhang A, Sun H, Wang X. Serum metabolomics as a novel diagnostic approach for disease: a systematic review. Anal Bioanal Chem. 2012 Sep;404(4):1239-45. doi: 10.1007/s00216-012-6117-1. Epub 2012 May 31. Review. PubMed PMID: 22648167. | yes | no | no | no | 1 | no | no | no | no | 0 | no | no | no | no | 0 | no | no | no | no | 0 |
| **177** | 121: Saligan LN, Kim HS. A systematic review of the association between immunogenomic markers and cancer-related fatigue. Brain Behav Immun. 2012 Aug;26(6):830-48. doi: 10.1016/j.bbi.2012.05.004. Epub 2012 May 14. Review. PubMed PMID: 22595751; PubMed Central PMCID: PMC3398196. | no | no | no | no | 0 | no | no | no | no | 0 | no | no | no | no | 0 | no | no | no | no | 0 |
| **178** | 122: Nieder C, Astner ST, Grosu AL. Glioblastoma research 2006-2010: pattern of citation and systematic review of highly cited articles. Clin Neurol Neurosurg. 2012 Nov;114(9):1207-10. doi: 10.1016/j.clineuro.2012.03.049. Epub 2012 Apr 18. Review. PubMed PMID: 22516416. | no | no | no | no | 0 | no | no | no | no | 0 | no | no | no | no | 0 | no | no | no | no | 0 |
| **179** | 123: Koetsier A, van der Veer SN, Jager KJ, Peek N, de Keizer NF. Control charts in healthcare quality improvement. A systematic review on adherence to methodological criteria. Methods Inf Med. 2012;51(3):189-98. doi: 10.3414/ME11-01-0055. Epub 2012 Apr 5. Review. PubMed PMID: 22476327. | no | no | no | no | 0 | no | no | no | no | 0 | no | no | no | no | 0 | no | no | no | no | 0 |
| **180** | 124: Sierink JC, Saltzherr TP, Reitsma JB, Van Delden OM, Luitse JS, Goslings JC. Systematic review and meta-analysis of immediate total-body computed tomography compared with selective radiological imaging of injured patients. Br J Surg. 2012 Jan;99 Suppl 1:52-8. doi: 10.1002/bjs.7760. Review. PubMed PMID: 22441856. | no | no | no | no | 0 | no | no | no | no | 0 | no | no | no | no | 0 | no | no | no | no | 0 |
| **181** | 125: Callesen AK, Mogensen O, Jensen AK, Kruse TA, Martinussen T, Jensen ON, Madsen JS. Reproducibility of mass spectrometry based protein profiles for diagnosis of ovarian cancer across clinical studies: A systematic review. J Proteomics. 2012 Jun 6;75(10):2758-72. doi: 10.1016/j.jprot.2012.02.007. Epub 2012 Feb 17. Review. PubMed PMID: 22366292. | no | yes | no | no | 1 | no | no | no | no | 0 | no | no | no | no | 0 | no | no | no | no | 0 |
| **182** | 126: Goldsmith L, Jackson L, O'Connor A, Skirton H. Direct-to-consumer genomic testing: systematic review of the literature on user perspectives. Eur J Hum Genet. 2012 Aug;20(8):811-6. doi: 10.1038/ejhg.2012.18. Epub 2012 Feb 15. Review. PubMed PMID: 22333900; PubMed Central PMCID: PMC3400732. | yes | yes | no | no | 2 | no | yes | no | no | 1 | no | no | no | no | 0 | no | no | no | no | 0 |
| **183** | 127: Bradley DT, Bourke TW, Fairley DJ, Borrow R, Shields MD, Young IS, Zipfel PF, Hughes AE. Genetic susceptibility to invasive meningococcal disease: MBL2 structural polymorphisms revisited in a large case-control study and a systematic review. Int J Immunogenet. 2012 Aug;39(4):328-37. doi: 10.1111/j.1744-313X.2012.01095.x. Epub 2012 Feb 2. Review. PubMed PMID: 22296677. | yes | yes | no | no | 2 | yes | yes | yes | yes | 4 | no | yes | yes | yes | 3 | yes | yes | yes | yes | 4 |
| **184** | 128: Johnson VA, Powell-Young YM, Torres ER, Spruill IJ. A systematic review of strategies that increase the recruitment and retention of African American adults in genetic and genomic studies. ABNF J. 2011 Winter;22(4):84-8. Review. PubMed PMID: 22165568; PubMed Central PMCID: PMC3439996. | no | yes | no | no | 1 | no | no | no | no | 0 | no | no | no | no | 0 | no | no | no | no | 0 |
| **185** | 129: Hugues JN. Impact of 'LH activity' supplementation on serum progesterone levels during controlled ovarian stimulation: a systematic review. Hum Reprod. 2012 Jan;27(1):232-43. doi: 10.1093/humrep/der380. Epub 2011 Nov 10. Review. PubMed PMID: 22081246. | no | no | no | no | 0 | no | no | no | no | 0 | no | no | no | no | 0 | no | no | no | no | 0 |
| **186** | 130: Veenendaal MV, van Abeelen AF, Painter RC, van der Post JA, Roseboom TJ. Consequences of hyperemesis gravidarum for offspring: a systematic review and meta-analysis. BJOG. 2011 Oct;118(11):1302-13. doi: 10.1111/j.1471-0528.2011.03023.x. Epub 2011 Jul 12. Review. PubMed PMID: 21749625. | no | no | no | no | 0 | no | no | no | no | 0 | no | no | no | no | 0 | no | no | no | no | 0 |
| **187** | 131: May KE, Villar J, Kirtley S, Kennedy SH, Becker CM. Endometrial alterations in endometriosis: a systematic review of putative biomarkers. Hum Reprod Update. 2011 Sep-Oct;17(5):637-53. doi: 10.1093/humupd/dmr013. Epub 2011 Jun 13. Review. PubMed PMID: 21672902. | no | no | no | no | 0 | no | no | no | no | 0 | no | no | no | no | 0 | no | no | no | no | 0 |
| **188** | 132: Schramm SJ, Mann GJ. Melanoma prognosis: a REMARK-based systematic review and bioinformatic analysis of immunohistochemical and gene microarray studies. Mol Cancer Ther. 2011 Aug;10(8):1520-8. doi: 10.1158/1535-7163.MCT-10-0901. Epub 2011 Jun 9. Review. PubMed PMID: 21659462. | yes | yes | yes | yes | 4 | yes | yes | yes | yes | 4 | yes | yes | yes | yes | 4 | yes | yes | yes | yes | 4 |
| **189** | 133: Musilová I, Kacerovský M, Tambor V, Tosner J. [Proteomics and biomarkers for detection of preterm labor: a systematic review]. Ceska Gynekol. 2011 Feb;76(1):37-45. Review. Czech. PubMed PMID: 21656999. | yes | yes | yes | yes | 4 | no | no | yes | no | 1 | no | no | no | no | 0 | no | no | no | no | 0 |
| **190** | 134: Al-Tarawneh SK, Border MB, Dibble CF, Bencharit S. Defining salivary biomarkers using mass spectrometry-based proteomics: a systematic review. OMICS. 2011 Jun;15(6):353-61. doi: 10.1089/omi.2010.0134. Epub 2011 May 13. Review. PubMed PMID: 21568728; PubMed Central PMCID: PMC3125555. | no | yes | no | yes | 2 | no | yes | yes | no | 2 | no | no | no | no | 0 | no | no | no | no | 0 |
| **191** | 135: Miller MH, Ferguson MA, Dillon JF. Systematic review of performance of non-invasive biomarkers in the evaluation of non-alcoholic fatty liver disease. Liver Int. 2011 Apr;31(4):461-73. doi: 10.1111/j.1478-3231.2011.02451.x. Epub 2011 Feb 1. Review. PubMed PMID: 21382157. | no | no | no | no | 0 | no | no | no | no | 0 | no | no | no | no | 0 | no | no | no | no | 0 |
| **192** | 136: Van der Sleen MI, Slot DE, Van Trijffel E, Winkel EG, Van der Weijden GA. Effectiveness of mechanical tongue cleaning on breath odour and tongue coating: a systematic review. Int J Dent Hyg. 2010 Nov;8(4):258-68. doi: 10.1111/j.1601-5037.2010.00479.x. Epub 2010 Sep 6. Review. PubMed PMID: 20961381. | no | no | no | no | 0 | no | no | no | no | 0 | no | no | no | no | 0 | no | no | no | no | 0 |
| **193** | 137: Wang H, Tso VK, Slupsky CM, Fedorak RN. Metabolomics and detection of colorectal cancer in humans: a systematic review. Future Oncol. 2010 Sep;6(9):1395-406. doi: 10.2217/fon.10.107. Review. PubMed PMID: 20919825. | yes | no | yes | no | 2 | no | no | no | no | 0 | no | no | no | no | 0 | no | no | no | no | 0 |
| **194** | 138: Liu Z, Ma Y, Yang J, Qin H. Upregulated and downregulated proteins in hepatocellular carcinoma: a systematic review of proteomic profiling studies. | no | yes | yes | yes | 3 | yes | yes | yes | no | 3 | no | yes | yes | no | 2 | yes | yes | yes | yes | 4 |
| | OMICS. 2011 Jan-Feb;15(1-2):61-71. doi: 10.1089/omi.2010.0061. Epub 2010 Aug 20. Review. PubMed PMID: 20726783. | | | | | | | | | | | | | | | | | | | | |
| **195** | 139: May KE, Conduit-Hulbert SA, Villar J, Kirtley S, Kennedy SH, Becker CM. Peripheral biomarkers of endometriosis: a systematic review. Hum Reprod Update. 2010 Nov-Dec;16(6):651-74. doi: 10.1093/humupd/dmq009. Epub 2010 May 12. Review. PubMed PMID: 20462942; PubMed Central PMCID: PMC2953938. | no | no | no | no | 0 | no | no | no | no | 0 | no | no | no | no | 0 | no | no | no | no | 0 |
| **196** | 140: Rew L, Mackert M, Bonevac D. A systematic review of literature about the genetic testing of adolescents. J Spec Pediatr Nurs. 2009 Oct;14(4):284-94. doi: 10.1111/j.1744-6155.2009.00210.x. Review. PubMed PMID: 19796327. | no | yes | no | no | 1 | no | no | no | no | 0 | no | no | no | no | 0 | no | no | no | no | 0 |
| **197** | 141: Lu JC, Coca SG, Patel UD, Cantley L, Parikh CR; Translational Research Investigating Biomarkers and Endpoints for Acute Kidney Injury (TRIBE-AKI) Consortium. Searching for genes that matter in acute kidney injury: a systematic review. Clin J Am Soc Nephrol. 2009 Jun;4(6):1020-31. doi: 10.2215/CJN.05411008. Epub 2009 May 14. Review. PubMed PMID: 19443624; PubMed Central PMCID: PMC2689876. | no | no | yes | no | 1 | no | no | no | no | 0 | no | no | no | no | 0 | no | no | no | no | 0 |
| **198** | 142: Atiomo W, Khalid S, Parameshweran S, Houda M, Layfield R. Proteomic biomarkers for the diagnosis and risk stratification of polycystic ovary syndrome: a systematic review. BJOG. 2009 Jan;116(2):137-43. doi: 10.1111/j.1471-0528.2008.02041.x. Review. PubMed PMID: 19076945. | no | yes | no | no | 1 | no | no | no | no | 0 | no | no | no | no | 0 | no | no | no | no | 0 |
| **199** | 143: Lichtenstein AH, Yetley EA, Lau J. Application of systematic review methodology to the field of nutrition. J Nutr. 2008 Dec;138(12):2297-306. doi: 10.3945/jn.108.097154. Review. PubMed PMID: 19022948; PubMed Central PMCID: PMC3415860. | no | no | no | no | 0 | no | no | no | no | 0 | no | no | no | no | 0 | no | no | no | no | 0 |
| **200** | 144: Yu YH, Kuo HK, Chang KW. The evolving transcriptome of head and neck squamous cell carcinoma: a systematic review. PLoS One. 2008 Sep 15;3(9):e3215. doi: 10.1371/journal.pone.0003215. PubMed PMID: 18791647; PubMed Central PMCID: PMC2533097. | yes | yes | yes | no | 3 | yes | yes | no | yes | 3 | yes | yes | yes | yes | 4 | yes | yes | yes | yes | 4 |
| **201** | 145: Di Bona D, Plaia A, Vasto S, Cavallone L, Lescai F, Franceschi C, Licastro F, Colonna-Romano G, Lio D, Candore G, Caruso C. Association between the interleukin-1beta polymorphisms and Alzheimer's disease: a systematic review and meta-analysis. Brain Res Rev. 2008 Nov;59(1):155-63. doi: | yes | yes | yes | no | 3 | yes | yes | yes | yes | 4 | no | no | no | yes | 1 | no | no | no | no | 0 |
| | 10.1016/j.brainresrev.2008.07.003. Epub 2008 Jul 21. Review. PubMed PMID: 18675847. | | | | | | | | | | | | | | | | | | | | |
| **202** | 146: Callesen AK, Vach W, Jørgensen PE, Cold S, Mogensen O, Kruse TA, Jensen ON, Madsen JS. Reproducibility of mass spectrometry based protein profiles for diagnosis of breast cancer across clinical studies: a systematic review. J Proteome Res. 2008 Apr;7(4):1395-402. doi: 10.1021/pr800115f. Epub 2008 Feb 28. Review. PubMed PMID: 18303834. | yes | yes | yes | no | 3 | yes | yes | no | no | 2 | no | no | no | no | 0 | no | no | no | no | 0 |
| **203** | 147: Subramonia-Iyer S, Sanderson S, Sagoo G, Higgins J, Burton H, Zimmern R, Kroese M, Brice P, Shaw-Smith C. Array-based comparative genomic hybridization for investigating chromosomal abnormalities in patients with learning disability: systematic review meta-analysis of diagnostic and false-positive yields. Genet Med. 2007 Feb;9(2):74-9. PubMed PMID: 17304048. | no | yes | no | no | 1 | no | no | no | no | 0 | no | no | no | no | 0 | no | no | no | no | 0 |
| **204** | 148: Leeflang MM, Cnossen JS, van der Post JA, Mol BW, Khan KS, ter Riet G. Accuracy of fibronectin tests for the prediction of pre-eclampsia: a systematic review. Eur J Obstet Gynecol Reprod Biol. 2007 Jul;133(1):12-9. Epub 2007 Feb 12. Review. PubMed PMID: 17293022. | no | no | no | no | 0 | no | no | no | no | 0 | no | no | no | no | 0 | no | no | no | no | 0 |
| **205** | 149: Adryan B, Teichmann SA. FlyTF: a systematic review of site-specific transcription factors in the fruit fly Drosophila melanogaster. Bioinformatics. 2006 Jun 15;22(12):1532-3. Epub 2006 Apr 13. PubMed PMID: 16613907. | yes | yes | yes | yes | 4 | yes | yes | yes | yes | 4 | no | no | no | no | 0 | no | no | no | no | 0 |
| **206** | 151: Panniers TL, Feuerbach RD, Soeken KL. Methods in informatics: using data derived from a systematic review of health care texts to develop a concept map for use in the neonatal intensive care setting. J Biomed Inform. 2003 Aug-Oct;36(4-5):232-9. Review. PubMed PMID: 14643718. | yes | yes | no | no | 2 | yes | yes | no | no | 2 | no | no | no | no | 0 | no | no | no | no | 0 |
| **207** | 152: Misra S, Crosby MA, Mungall CJ, Matthews BB, Campbell KS, Hradecky P, Huang Y, Kaminker JS, Millburn GH, Prochnik SE, Smith CD, Tupy JL, Whitfied EJ, Bayraktaroglu L, Berman BP, Bettencourt BR, Celniker SE, de Grey AD, Drysdale RA, Harris NL, Richter J, Russo S, Schroeder AJ, Shu SQ, Stapleton M, Yamada C, Ashburner M, Gelbart WM, Rubin GM, Lewis SE. Annotation of the Drosophila melanogaster euchromatic genome: a systematic review. Genome Biol. 2002;3(12):RESEARCH0083. Epub 2002 Dec 31. Review. PubMed PMID: 12537572; PubMed Central PMCID: PMC151185. | no | yes | yes | yes | 3 | yes | yes | yes | yes | 4 | no | no | no | no | 0 | no | no | no | no | 0 |

<table>
<tr><td>Number of references approved by steps:</td><td>55</td><td>72</td><td>24</td><td>34</td><td>59</td><td>32</td><td>34</td><td>25</td><td>32</td><td>33</td><td>10</td><td>12</td><td>9</td><td>9</td><td>12</td><td>9</td><td>9</td><td>8</td><td>8</td><td>9</td></tr>
<tr><td rowspan="2">References screened: 207</td><td colspan="4">Eliminated:</td><td>148</td><td colspan="4">Eliminated:</td><td>26</td><td colspan="4">Eliminated:</td><td>21</td><td colspan="4">Eliminated:</td><td>3</td></tr>
<tr><td colspan="19">Eliminated (score evaluation):</td><td>1</td></tr>
<tr><td colspan="21">References included in the SLR: 8</td></tr>
</table>